\documentclass[aps,prb,superscriptaddress,longbibliography,twocolumn,showkeys]{revtex4-2}
\usepackage{bm}
\usepackage{lmodern}

\usepackage[utf8]{inputenc}
\usepackage{amsmath,bm}
\usepackage{amssymb}
\usepackage{graphicx}
\usepackage{epstopdf}
\usepackage{xcolor}
\usepackage{enumerate}
\usepackage[normalem]{ulem}
\usepackage[english]{babel}
\usepackage{braket}
\usepackage{natbib}
\usepackage{float}
\usepackage{stmaryrd}
\usepackage{hyperref}
\usepackage{physics}
\usepackage{upgreek}
\usepackage[T1]{fontenc}
\usepackage{textcomp}

\fontfamily{Calibri}\selectfont
\begin{document}

\title{Single-electron occupation in quantum dot arrays at selectable plunger gate voltage}

\author{Marcel Meyer}
\author{Corentin D\'{e}prez}
\author{Ilja N. Meijer}
\author{Florian K. Unseld}
\affiliation{QuTech and Kavli Institute of Nanoscience, Delft University of Technology, PO Box 5046, 2600 GA Delft, The Netherlands}
\author{Saurabh Karwal}
\author{Amir Sammak}
\affiliation{QuTech and Netherlands Organisation for Applied Scientific Research (TNO), PO Box 155, 2600 AD Delft, The Netherlands}
\author{Giordano Scappucci}
\author{Lieven M. K. Vandersypen}
\author{Menno Veldhorst}
\email{corresponding author: m.veldhorst@tudelft.nl}
\affiliation{QuTech and Kavli Institute of Nanoscience, Delft University of Technology, PO Box 5046, 2600 GA Delft, The Netherlands}

\date{\today}

\begin{abstract}
The small footprint of semiconductor qubits is favorable for scalable quantum computing. However, their size also makes them sensitive to their local environment and variations in gate structure. Currently, each device requires tailored gate voltages to confine a single charge per quantum dot, clearly challenging scalability. Here, we tune these gate voltages and equalize them solely through the temporary application of stress voltages. In a double quantum dot, we reach a stable (1,1) charge state at identical and predetermined plunger gate voltage and for various interdot couplings. Applying our findings, we tune a 2$\times$2 quadruple quantum dot such that the (1,1,1,1) charge state is reached when all plunger gates are set to 1 V. The ability to define required gate voltages may relax requirements on control electronics and operations for spin qubit devices, providing means to advance quantum hardware.
\end{abstract}

\keywords{Quantum Dot, Single-electron Occupation, Uniformity, Stress Voltage, Spin Qubit}

\maketitle


\section*{Introduction}
Semiconductor spin qubits have become a compelling platform for quantum computation. Single qubit gate fidelities of 99.99\%~\cite{Lawrie2023} and two-qubit gate fidelities exceeding 99\%~\cite{Madzik2022,Noiri2022,Xue2022,Mills2022} have been demonstrated. A moderate sensitivity to thermal effects allowed for the implementation of quantum operations above one Kelvin~\cite{Petit2020,Yang2020,Camenzind2022}. Furthermore, the small size of semiconductor spin qubits and their compatibility with advanced semiconductor manufacturing~\cite{Bourdet2018,Ansaloni2020,Zwerver2022} may facilitate devices with large numbers of qubits as required for practical applications. Recent advances in the material platforms supported the realization of a $2\times2$ qubit array in germanium~\cite{Hendrickx2021}, a linear six qubit system in silicon~\cite{Philips2022}, and the operation of a 16 quantum dot crossbar array~\cite{Borsoi2023}. However, scaling up the number of qubits is challenging, especially when considering the numbers needed for fault-tolerant quantum computation~\cite{Fowler2012,Wecker2014,Terhal2015}. A particular challenge lies in the sensitivity of qubits to their environment leading to considerable variations of their properties, a notion that was already highlighted in the seminal work on quantum computation by Loss and DiVincenzo \cite{LossDiVincenzo1998}.

Substantial reductions in variability have been achieved through progress in heterostructure growth and device fabrication. For instance, these efforts focus on reducing material disorder~\cite{Schäffler1997,Borselli2011,Mi2015,Li2015,Esposti2022,Wuetz2023,Stehouwer2023,Myronov2023}, advancing device fabrication~\cite{Dodson2020,Lawrie2020,Ha2022} and addressing fluctuations in mechanical stress induced by the deposition of metallic gate electrodes~\cite{Thorbeck2015,Park2016,Stein2021}. However, significant variations remain observable in current devices~\cite{Zajac2016,Mills2019,Borsoi2023} and it is an open question whether sufficient uniformity can be reached through material development alone.

Alternatively, fluctuations in the potential landscape can be compensated by temporarily applying stress voltages~\cite{Huang2014,Laroche2015,Su2019,Meyer2023}. An alternating sequence of stress voltages and pinch-off measurements has already enabled on-demand reshaping of pinch-off voltage characteristics and their homogenization without signs of reduced device stability afterwards. Furthermore, such sequences allowed to alter the potential offset of a single electron transistor (SET) at a temperature of $\approx 4.2~$K~\cite{Meyer2023}. Yet, this methodology has not been applied to individual electrons in a quantum dot. Also, overcoming qubit variations in quantum processors will require the tuning of multiple quantum dots.

Here, we demonstrate the use of stress voltages to tune the potential landscape in a quantum dot array. We show that this approach allows to change and equalize the plunger gate voltages required to reach single-electron occupation in a double quantum dot without changing any other gate voltages. Importantly, we find that the resulting confining potential remains stable for hours afterwards. To illustrate its robustness and versatility, we demonstrate that the method employed can be applied at various barrier voltages and thus interdot tunnel couplings. Furthermore, we show that the procedure can be extended to homogenize the plunger gate voltages defining the single occupation charge state in a $2\times2$ quantum dot system.

\section*{Results}

\begin{figure}[]
\centering
\includegraphics[]{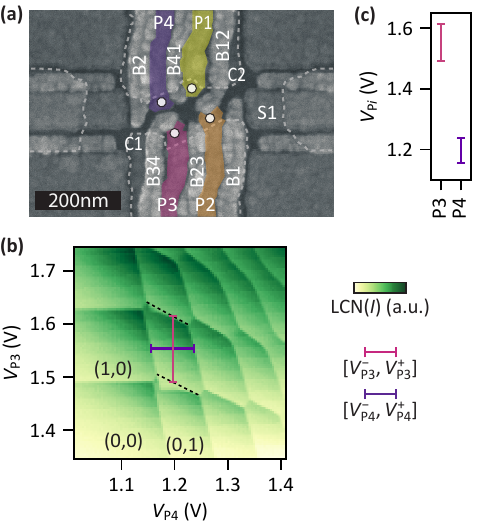}
\caption{\textbf{Device and tuning of a double quantum dot.} \textbf{(a)} Scanning electron micrograph of a device nominally identical to the one under study. Confinement (C$i$) and barrier (B$i$ and B$ij$) gates are designed to define four quantum dots indicated by the white circles. Their charge occupation is controlled by four plunger (P$i$) gates. Confinement gates are outlined by dashed lines for clarity. A sensor quantum dot is formed under S1 and measured in transport. \textbf{(b)} Charge stability diagram showing the single-electron occupation of the Q3-Q4 double quantum dot formed underneath P3 and P4. The plotted signal is locally contrast normalized (LCN) to increase the visibility of the charge transition lines as described in the methods section. Dashed lines connect charge triple degeneracy points and thereby indicate transitions of the charge ground state which cannot be observed directly due to latching effects. The plunger gate voltage ranges $[V_{{\rm P}i}^{-},V_{{\rm P}i}^{+}]$ that set a $(1,1)$ charge state are indicated by vertical and horizontal bars. The ranges are extracted around the center point of the (1,1) charge region (see methods). Unprocessed data shown in supplementary section~\ref{sup-sec:raw-fig1-4}. \textbf{(c)} Plunger gate voltage ranges $[V_{{\rm P}i}^{-},V_{{\rm P}i}^{+}]$ as extracted in \textbf{(b)}.}
\label{fig:Fig1}
\end{figure}

Fig.~\ref{fig:Fig1}.a shows a scanning electron micrograph of a device nominally identical to the one under study in this work, which is fabricated on a $^{28}$Si/SiGe heterostructure~\cite{Esposti2023} (see methods). The gate design allows for the formation of a $2\times2$ quantum dot array (white circles) and two adjacent single electron transistors (SETs) on the left and right side~\cite{Unseld2023}. We form the quantum dots Q3 and Q4 underneath the plunger gates P3 and P4 and also tune up the SET below the sensor gate S1. The left side of the device is operated as an electron reservoir. Fig.~\ref{fig:Fig1}.b depicts a charge stability diagram recorded after the initial tuning. It shows the typical honeycomb pattern of a double quantum dot and depletion down to the $(N_3,N_4)=(1,1)$ charge state with $N_i$ the charge occupation of Q$i$.

The charge stability diagram reveals a large asymmetry in the plunger gate voltages required to reach the single-electron regime. The voltage ranges $[V_{{\rm P}i}^{-},V_{{\rm P}i}^{+}]$ from the first to the second charge transition line of the two quantum dots are indicated by a horizontal and a vertical bar (see methods for the definition). As illustrated in Fig.~\ref{fig:Fig1}.c those ranges do not overlap for the two quantum dots and in particular we find a separation of more than 2(4) times the Q3(Q4) charging voltage $V_{{\rm P}i}^{\rm C} = V_{{\rm P}i}^{+}-V_{{\rm P}i}^{-}$. While this is a rather extreme case, significant asymmetries of the plunger gate voltage ranges loading a single electron are commonly observed in quantum dot devices~\cite{Zajac2016,Noiri2022b,Wuetz2022,Ziegler2023, Borsoi2023}. Therefore, if single-electron occupation can be achieved at equal plunger gate voltages in the device of Fig.~\ref{fig:Fig1} this would provide good prospects for the homogenization of the required plunger gate voltages in other devices that already are intrinsically more uniform.

\begin{figure*}[]
\centering
\includegraphics[]{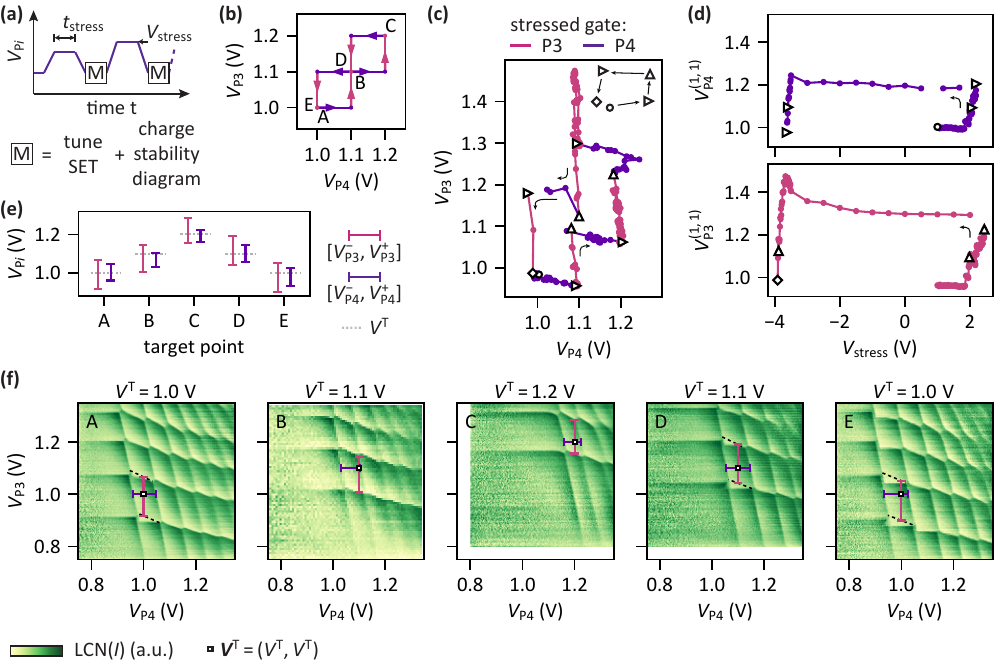}
\caption{\textbf{Single-electron occupation at predetermined plunger gate voltages through voltage stressing.} \textbf{(a)}~Schematic of the stress-measure sequence applied to shift the voltages required to obtain the $(1,1)$ charge state. Increasing stress voltages $V_{\rm stress}$ are applied for $t_{\rm stress}=1~$min interleaved by charge stability diagram measurements. \textbf{(b)}~Expected trajectory for the center of the (1,1) charge region $\textbf{\textit{V}}^{(1,1)}$ in the ($V_{\rm P3}$,$V_{\rm P4}$) plane during the tuning procedure as defined prior to conducting the experiment. The color of the path refers to the plunger gate being stressed. \textbf{(c)}~Actual trajectory of $\textbf{\textit{V}}^{(1,1)}$ followed during the tuning procedure. The triangle, circles and diamond mark the starting point, (intermediate) targets and the endpoint of the path, respectively. Black arrows indicate the time flow. \textbf{(d)}~$V^{(1,1)}_{\rm P3}$ (bottom) and $V^{(1,1)}_{\rm P4}$ (top) as a function of the applied stress voltage $V_{\rm stress}$. The triangle, circles and diamond mark the same points as in \textbf{(c)} and black arrows indicate the time flow. \textbf{(e)}~Plunger gate voltage ranges $[V_{{\rm P}i}^{-},V_{{\rm P}i}^{+}]$ that keep the double quantum dot in the $(1,1)$ charge state after tuning (see methods). Targets are indicated by the dotted lines. \textbf{(f)}~Corresponding charge stability diagrams recorded after the application of the respective stress voltage sequences. The white square markers show the target voltages $\textbf{\textit{V}}^{\rm T}=(V^{\rm T},V^{\rm T})$. Plunger gate voltage ranges $[V_{{\rm P}i}^{-},V_{{\rm P}i}^{+}]$ that keep the system in the $(1,1)$ charge state are indicated by vertical and horizontal bars. Dashed lines indicate transitions of the charge ground state which cannot be observed directly due to latching effects. Unprocessed data shown in supplementary section~\ref{sup-sec:raw-fig1-4}.}
\label{fig:Fig2}
\end{figure*}

\subsection*{(1,1) charge occupation at predetermined plunger gate voltage}
To increase the potential uniformity, we follow our previous work \cite{Meyer2023} and apply stress voltages $V_{\rm stress}$ on gate electrodes to reshape the background potential landscape. We aim to tune the system such that the (1,1) charge state is reached at predetermined plunger gate voltage. Specifically we target to load a single electron per quantum dot for $V_{\rm P3}=V_{\rm P4}=V^{\rm T}$ with $V^{\rm T}=1~$V, 1.1~V and 1.2~V by sequentially tuning the potential below the two plunger gates following the path shown in Fig.~\ref{fig:Fig2}.b. Fig.~\ref{fig:Fig2}.a illustrates the employed procedure for a single plunger gate P$i$. We apply a stress voltage $V_{\rm stress}$ for $t_{\rm stress}=1~$min. Afterwards, we measure charge stability diagrams around $V_{{\rm P}i}=V^{\rm T}$ and if necessary the sensor gate voltage $V_{\rm S1}$ is compensated to restore maximum sensitivity of the SET. From the charge stability diagrams we then extract the voltage range $[V_{{\rm P}i}^{-},V_{{\rm P}i}^{+}]$ required to reach single charge occupation. If setting the target voltage does not yield the targeted electron occupation in Q$i$ ($V^{\rm T}$ not in $[V_{{\rm P}i}^{-},V_{{\rm P}i}^{+}]$) the sequence is repeated with an increased (decreased) stress voltage to shift the voltage range further upward (downward). If a single electron is loaded at the target voltage configuration we stop applying stress voltages to P$i$ and analogously tune the potential of the other quantum dot. After the initial tune up (Fig.~\ref{fig:Fig1}), we first follow the stressing procedure to lower the required plunger gate voltage ranges $[V_{{\rm P}i}^{-},V_{{\rm P}i}^{+}]$ to reach single-electron occupancy at 1~V. During this process we adjust the barrier gate B2 voltage in order to maintain a significant tunnel rate. Then, we perform the stressing experiment and advance from point A to E in Fig.~\ref{fig:Fig2}.b. Here, we only change the sensor gate S1 voltage and keep all other gate voltages constant (see supplementary section~\ref{sup-sec:voltage-configs} for the voltage settings).

Fig.~\ref{fig:Fig2}.f shows charge stability diagrams recorded after tuning toward the predefined targets $V^{\rm T}$. A clear shift of the (1,1) charge region to higher plunger gate voltages and then back down is observable. Furthermore, after the completion of each tuning, setting the plunger gate voltages $(V_{\rm P3}, V_{\rm P4})$ to $\textbf{\textit{V}}^{\rm T}=(V^{\rm T},V^{\rm T})$ (white square marker) loads a single electron per quantum dot as also highlighted in Fig.~\ref{fig:Fig2}.e showing the extracted voltage ranges $[V_{{\rm P}i}^{-},V_{{\rm P}i}^{+}]$. This demonstrates tunability of the chemical potentials and control over the electron occupation in a double quantum dot through the temporary application of stress voltage. Note that charge latching is reduced (increased) when tuning the voltage ranges $[V_{{\rm P}i}^{-},V_{{\rm P}i}^{+}]$ upwards (downwards). This suggests a crosstalk effect of the applied stress voltages on the surrounding tunnel barrier potentials.

Fig.~\ref{fig:Fig2}.c shows the reconstructed evolution of the center point of the (1,1) charge region $\textbf{\textit{V}}^\mathrm{(1,1)}=(V^{(1,1)}_{\rm P3},V^{(1,1)}_{\rm P4})$ during the tuning procedure (see methods section). Overall, the experimental trajectory reproduces qualitatively the intended one shown in Fig.~\ref{fig:Fig2}.b. The predominantly horizontal and vertical progressions in the $(V^{(1,1)}_{\rm P3},V^{(1,1)}_{\rm P4})$ plane suggest limited crosstalk, i.e. applying stress voltages to one gate P$i$ only has a small effect on the charge transition voltages of the quantum dot below the other plunger gate. Quantitatively, we find slopes $dV^{(1,1)}_{{\rm P}i}/dV^{(1,1)}_{{\rm P}j}$ between $-0.31~$V/V and $-0.04~$V/V. The sign of these slopes is consistent with the sign of the capacitive shift of the transition line voltage of Q$j$ when the plunger gate voltage $V_{{\rm P}i}$ is changed (see supplementary section~\ref{sup-sec:crosstalk}). Correcting for this effect, we obtain the change of the charge transition voltages of Q$j$ induced exclusively by the application of stress voltages set to P$i$. We find crosstalks of $(+0.37\pm0.03)~$V/V and $(+0.19\pm0.03)~$V/V for P3 on Q4 and P4 on Q3 respectively. Overall, while these crosstalk effects could be compensated for, the simple approach presented here allowed to tune the potentials of the quantum dots to the predetermined targets.

In Fig.~\ref{fig:Fig2}.d the center voltages $V^{(1,1)}_{3}$ and $V^{(1,1)}_{4}$ are plotted as a function of the applied stress voltage $V_{\rm stress}$. We recover the typical hysteresis cycle observed when tuning pinch-off voltages using an analogous method in similar devices~\cite{Meyer2023}. Noticeably, for steadily decreasing stress voltages there is an initial increase in $V^{(1,1)}_{{\rm P}i}$ before it rapidly drops to lower voltages at $V_{\rm stress}\approx-4~$V. In Fig.~\ref{fig:Fig2}.c this manifests as non-monotonic progressions of $\textbf{\textit{V}}^{(1,1)}$ between the target points C and D. $V^{(1,1)}_{\rm P4}$ and $V^{(1,1)}_{\rm P3}$ initially increase by 40~mV and 180~mV, respectively, before they decrease and approach $V^{\rm T}=1.1~$V.

Summarizing, Fig.~\ref{fig:Fig2} demonstrates that the background potential in the quantum well can be reshaped such that each quantum dot can be occupied with one electron using uniform plunger gate voltages.

\begin{figure}[]
\centering
\includegraphics[]{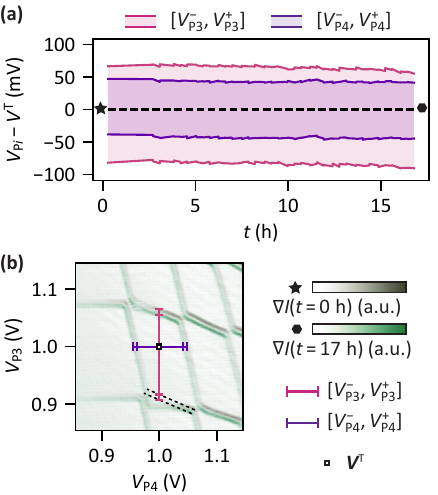}
\caption{\textbf{Stability of the (1,1) charge state after stress tuning.} \textbf{(a)} Time traces of the plunger gate voltage ranges that keep the system in the $(1,1)$ charge state (see methods for the definition) after the application of a sequence of increasing stress voltages. $t$ is the time after the application of the last stress voltage. Note that the underlying charge stability diagram measurements were interleaved with charge noise measurements on the sensor (see supplementary section~\ref{sup-sec:charge-noise}). Additional traces are presented in supplementary section~\ref{sup-sec:time-traces}. \textbf{(b)} Overlay of charge stability diagrams taken at the beginning (star, olive green) and end (hexagon, light green) of the time trace shown in \textbf{(a)}. Horizontal and vertical bars indicate the respective plunger gate voltage ranges that keep the system in the (1,1) charge state. Dashed lines indicate transitions of the charge ground state which cannot be observed directly due to latching effects. Unprocessed data shown in supplementary section~\ref{sup-sec:raw-fig1-4}.}
\label{fig:Fig3}
\end{figure}

\subsection*{Time stability}
To understand the impact of stress voltages on device stability, we record multiple charge stability diagrams as a function of time after the initial stress tuning towards $V^{\rm T}=1~$V (A in Fig.\ref{fig:Fig2}.d). Fig.~\ref{fig:Fig3}.a shows the extracted evolution of the plunger gate voltage range that keeps the quantum dots Q3 and Q4 in the single-electron occupation. Here, the time $t$ refers to the time since the last application of a stress voltage and voltages are plotted relative to $V^{\rm T}$. We find that the double quantum dot system remains in the $(1,1)$ charge state for more than 15~h showing only a weak drift. This is confirmed by standard deviations of 3~mV, 3~mV, 2~mV, and 1~mV for $V_{\rm P3}^{-}$, $V_{\rm P3}^{+}$, $V_{\rm P4}^{-}$, and $V_{\rm P4}^{+}$, respectively, which remain negligible compared to the charging voltages of 148~mV and 87~mV for Q3 and Q4, respectively. Overlaying the charge stability diagrams recorded at $t=0~$h and at $t=17~$h, as depicted in Fig.~\ref{fig:Fig3}.b, provides further confirmation of the device stability. Additional time traces demonstrating stability up to 40~h after the application of the last stress voltages are presented in supplementary section~\ref{sup-sec:time-traces}. Moreover, we find no increase in the charge noise sensed by the right SET when comparing to typical values for such devices (see supplementary section~\ref{sup-sec:charge-noise}). Note that the charge noise amplitude measured by the SET might differ from the charge noise level that would affect the coherence of qubits in the array. Nevertheless, we conclude that there are no signs of decreased device stability caused by the application of stress voltages.

\begin{figure*}[]
\centering
\includegraphics[]{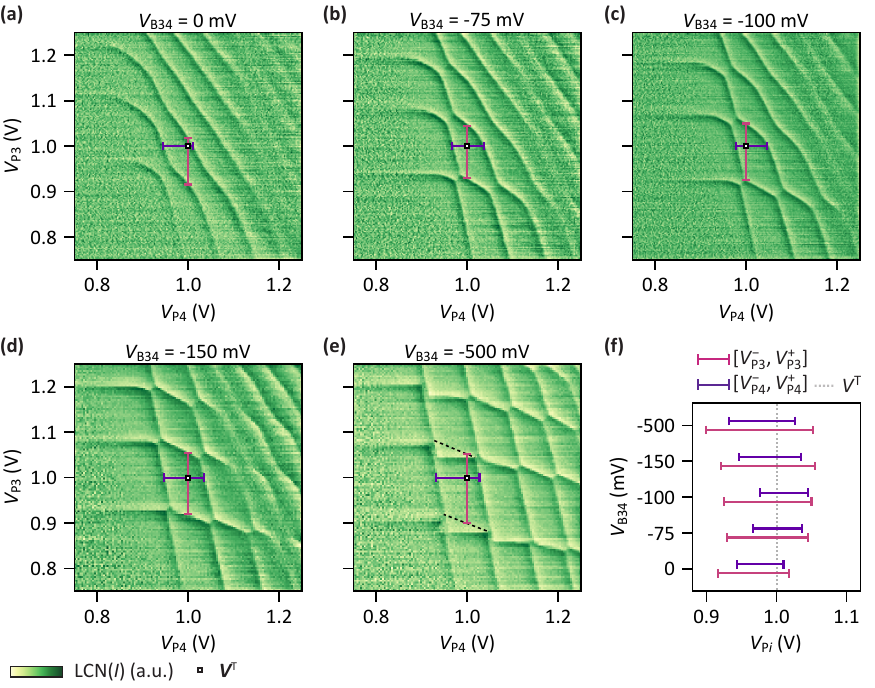}
\caption{\textbf{Single-electron occupation at predetermined plunger gate voltage for high and low interdot coupling.} \textbf{(a)-(e)}~Charge stability diagrams measured after tuning the system through applying stress voltages such that the (1,1) charge state is the ground state when applying the plunger gate voltages $\textbf{\textit{V}}^{\rm T}=(1~{\rm V},1~{\rm V})$ (white square marker). In each case a different barrier gate voltage $V_{\rm B34}$ is set before the tuning (labelled in the plot titles). The range of plunger gate voltages $[V_{{\rm P}i}^{-},V_{{\rm P}i}^{+}]$ that keep the system in the $(1,1)$ charge state is indicated by horizontal and vertical bars (see methods). Dashed lines indicate transitions of the charge ground state which cannot be observed directly due to latching effects. The unprocessed data is shown in supplementary section~\ref{sup-sec:raw-fig1-4}. \textbf{(f)} Plunger gate voltage ranges $[V_{{\rm P}i}^{-},V_{{\rm P}i}^{+}]$ extracted from \textbf{(a)-(e)}. The dotted line indicates the target voltage $V^{\rm T}=1~$V.}
\label{fig:Fig4}
\end{figure*}

\subsection*{Predetermined plunger gate voltage for tunnel coupled quantum dots}
We now address the question whether single-electron occupation can still be achieved by a predetermined gate voltage, when changing the coupling between the quantum dots. In our double quantum dot system, we can control the interdot coupling by adjusting the barrier gate B34 voltage to tune the system from strong to weak coupling quantum dots. We achieve this by varying the barrier gate voltages between $0~$V and $-0.5~$V. After setting a barrier gate voltage, we apply stress voltages to the plunger gates to obtain the $(1,1)$ charge state at $\textbf{\textit{V}}^{\rm T}=(1~{\rm V},1~{\rm V})$. Fig.~\ref{fig:Fig4}.a-e shows the resulting charge stability diagrams. The charge transition line pattern changes from exhibiting nearly diagonal lines at $V_{\rm B34}=0~$mV towards a rectangular grid-like pattern at $V_{\rm B34}=-500~$mV, revealing the transition from high to low coupling. In all cases the application of stress voltage sequences allows to obtain the $(1,1)$ charge state at $\textbf{\textit{V}}^{\rm T}=(1~{\rm V},1~{\rm V})$. This is confirmed by the extracted voltage ranges $[V_{{\rm P}i}^{-},V_{{\rm P}i}^{+}]$ plotted in Fig.~\ref{fig:Fig4}.f. Crucially, this is achieved without defining virtual gates. We conclude that for a wide range of interdot couplings single-electron occupation can be achieved at predetermined plunger gate voltage independently of the applied barrier voltage. 

\begin{figure}[]
\centering
\includegraphics[]{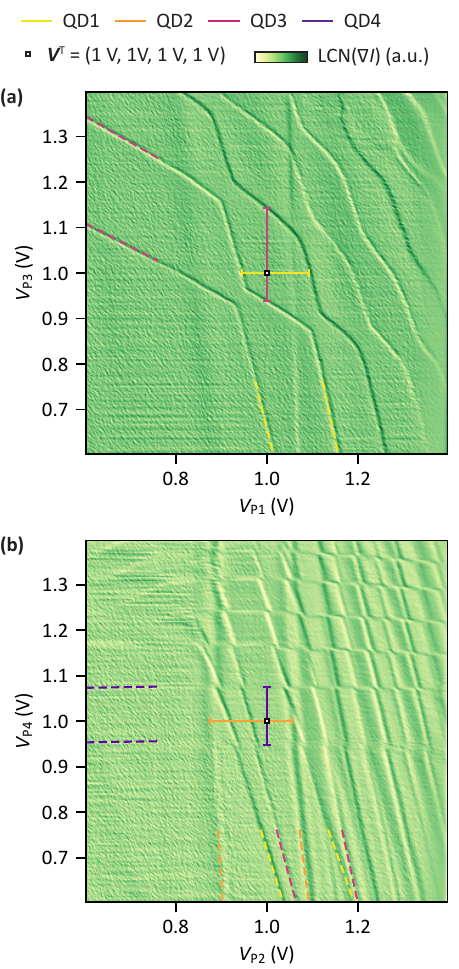}
\caption{\textbf{(1,1,1,1) charge state at 1~V on all plunger gates} \textbf{(a), (b)} Charge stability diagrams recorded after applying stress voltage sequences to tune the (1,1,1,1) charge state to be the ground state when all plunger gate voltages are set to 1~V. The first two transition lines of each quantum dot are indicated by dashed lines. The voltage ranges to keep the system in the (1,1,1,1) charge state are indicated by horizontal and vertical bars (see methods). A white square marks the point when all plunger gates are at 1~V. The plotted signal is the summation of several charge stability diagrams with identical voltage ranges recorded for slightly varied voltages on the SET plunger S1 (see supplementary section~\ref{sup-sec:raw-fig5}). Contrast is enhanced by a local contrast normalization (LCN). \textbf{(a)} shows charge transitions of Q1 and Q3 and \textbf{(b)} exhibits charge transition lines of all four dots.}
\label{fig:Fig5}
\end{figure}

\subsection*{(1,1,1,1) charge state at (1,1,1,1)~V}
Finally, we utilize our findings to tune a $2\times2$ quantum dot array such that the $(N_1,N_2,N_3,N_4)=(1,1,1,1)$ charge state is the ground state when all plunger gate voltages are set to 1~V. Starting from the Q3-Q4 double quantum dot, we form the quantum dots Q1 and Q2 which are predominantly controlled by the plunger gates P1 and P2. Then, the system is tuned solely through tailored stress voltage sequences applied to the plunger gates. Fig.~\ref{fig:Fig5} shows two charge stability diagrams recorded after this tuning process unveiling four sets of charge transition lines. These can be associated with the four quantum dots by analysing further charge stability diagrams recorded by sweeping additional plunger gate combinations (see supplementary section~\ref{sup-sec:four-QD-identification}). Yellow, orange, red and purple dashed lines mark the first two charge addition voltages of quantum dot Q1, Q2, Q3 and Q4, respectively. The target voltage configuration $\textbf{\textit{V}}^{\rm T}=(V^{\rm T}_{\rm P1},V^{\rm T}_{\rm P2},V^{\rm T}_{\rm P3},V^{\rm T}_{\rm P4})=(1~{\rm V},1~{\rm V},1~{\rm V},1~{\rm V})$ is shown by a white square marker and the voltage ranges that keep the system in the (1,1,1,1) charge state are indicated by horizontal and vertical bars. $\textbf{\textit{V}}^{\rm T}$ clearly falls between the first two charge transition lines for all four quantum dots confirming that we reached the targeted configuration. Note that all quantum dots are strongly affected by plunger gate P2 and P4 as observable in Fig.~\ref{fig:Fig5}.b. However, in Fig.~\ref{fig:Fig5}.a the voltages on P1 and P3 only seem to affect the charge occupation of Q1 and Q3. We speculate this behavior to originate from asymmetries in the gate layout and device imperfections~\cite{Unseld2023}. Crucially, we find that the stressing procedure is effective for the tuning of a nonlinear quadruple quantum dot array.

\section*{Discussion}

In summary, we have shown that single-electron occupation in quantum dots can be achieved at equal predetermined plunger gate voltage, by making use of a stress-voltage based procedure. Importantly, we find that after such a tuning the systems remains stable for hours only exhibiting small progressive drifts which do not affect the charge configuration. We envision that the stressing methodology may find several applications in semiconductor quantum technology. For instance, it may facilitate the operation of crossbar arrays which crucially rely on shared gate voltages~\cite{Li2018,Borsoi2023}. While our experiments suggest tunability of the entire potential landscape, more research is needed to understand the level of control over the barrier potentials. A predetermined gate voltage to set a given charge state can also relax the requirements on control electronics and facilitate their integration. Furthermore, we envision that stressing voltages can provide tunability of other parameters. For example, the $g$-tensor of germanium qubits is strongly dependent on the electric field~\cite{Lawrie2020, Hendrickx2023}, such that stressing voltages may provide tunability over the qubit resonance frequency. We therefore envision that stressing procedures may become a standard and essential routine in the tuning of large quantum circuits.

\section*{Material and methods}

\subsection*{Heterostructure and device fabrication}
The device under study in this work is fabricated on a $^{28}$Si/SiGe heterostructure~\cite{Esposti2023} which is based on a Si wafer. First, a linearly graded Si$_{1-x}$Ge$_{x}$ buffer with $x$ varying from 0 to 0.3 is grown followed by a 300~nm relaxed Si$_{0.7}$Ge$_{0.3}$ layer. A 7~nm purified (800~ppm) $^{28}\text{Si}$ layer defines the quantum well and is separated from the gate stack by another 30~nm thick relaxed Si$_{0.7}$Ge$_{0.3}$ buffer that is passivated in dichlorosilane at 500~$^\circ$C. Phosphorus ion implantation is utilized to contact the two dimensional electron gas and a 10~nm aluminum oxide layer precedes the deposition of gate electrodes. The latter are spread across three layers and made of Ti/Pd deposited via electron beam evaporation. They are separated by 5~nm thick layers of aluminium oxide. In all cases aluminium oxide is deposited via atomic layer deposition~\cite{Lawrie2020}.

\subsection*{Setup and voltage pulses}
All measurements are performed in a dilution refrigerator at a base temperature of $\approx 20~$mK. The gate voltages are supplied by digital analog converters (DACs) with a resolution of 18~bit and a voltage range of $\pm 4~$V which was amplified to $\pm 20~$V for the plunger gates. The current through the SET is measured via a current-to-voltage converter connected to a digitizer module. Confinement and stress voltages are applied via the DACs while charge stability diagrams are recorded by sending fast voltage pulses. The latter are generated by an arbitrary waveform generator (AWG). DAC and AWG voltage signals are merged with a bias tee located on the sample PCB at the mixing chamber stage. AWG pulses are modified to correct for voltage drifts caused by (dis)charging of the bias tees. Furthermore, cross-capacitive shifts from P3 and P4 on the sensing dot potential are compensated for by proportionally adjusting $V_{\rm S1}$ when sweeping the plunger gate voltages $V_{{\rm P}i}$ ($\Delta V_{\rm S1} / \Delta V_{{\rm P}i} < 0.01$).

\subsection*{Local contrast normalization}
In voltage scans spanning a large range, cross-capacitive coupling of the plunger gates to the SET can cause significant variations in sensor sensitivity. This leads to contrast fluctuations across the charge stability diagram and hampers identification of charge transition lines. We compensated for this effect by applying a local contrast normalization (LCN). In essence, a smoothed  charge stability map is subtracted to compensate for a slowly varying offset after which a smoothed local variance is utilized to locally normalize the signal:
\begin{equation}
    \nonumber
    {\rm LCN}(I) = \frac{I - I \ast f_{\mathrm{Gaussian}}}{\sqrt{(I - I \ast f_{\mathrm{Gaussian}})^2 \ast f_{\mathrm{Gaussian}}}}
\end{equation}
Here, the asterisk denominates a convolution, $I$ is the sensor signal and $f_{\rm Gaussian}$ refers to a normal distribution with a mean and variance chosen between 4 and 50~pixels.

\subsection*{Extraction of characteristic voltages from charge stability diagrams}
For each charge stability diagram we identify the coordinates of the charge triple degeneracy points (triple points) that constitute the corners of the (1,1) charge region. From these we calculate the voltage ranges $[V_{{\rm P}i}^{-},V_{{\rm P}i}^{+}]$ that keep the system in the (1,1) charge state around the center point $\textbf{\textit{V}}^{(1,1)}$ (in Fig.~\ref{fig:Fig1}) or the target voltages $\textbf{\textit{V}}^{\rm T}$ (in all other figures). The center point $\textbf{\textit{V}}^{(1,1)}$ of the (1,1) charge region is determined as the centroid of the triple points at the $(2,0)-(1,1)$ and $(1,1)-(2,0)$ charge transitions. Note that the voltage ranges $[V_{{\rm P}i}^{-},V_{{\rm P}i}^{+}]$ are a measure of the maximum voltage variation on a single plunger gate for which the charge state remains constant. When taking into account more than a single gate voltage a polytope describes the applicable gate voltages that keep the charge state at single electron occupation. For instance, when considering two plunger gates the polytope would be the hexagon typically found in a double quantum dot honeycomb pattern. While we utilize one-dimensional voltage ranges $[V_{{\rm P}i}^{-},V_{{\rm P}i}^{+}]$ to ease visualizations, after all stressing experiments the target voltage point $\textbf{\textit{V}}^{\rm T}$ lies inside the single charge occupation region (inside the respective gate voltage polytope).

We have used the triple points for the analysis because of their robustness against latching effects. For instance, in Fig.~\ref{fig:Fig1}.b the dashed lines show reconstructed charge transition lines of quantum dot Q3 which has a weak coupling to the nearby charge reservoir. Consequentially, $[V_{{\rm P}i}^{-},V_{{\rm P}i}^{+}]$ can include regions of meta-stable charge state (in between the observed and the reconstructed charge transition). This does not impact our conclusions because, at the end of all stressing experiments, the target voltage point $\textbf{\textit{V}}^{\rm T}$ lies in a region of stable charge state.

\section*{Data availability}
The data and analysis supporting this work are openly available in a public Zenodo repository at \url{https://doi.org/10.5281/zenodo.8322422}~\cite{ZENODO_DATA}.

\section*{Acknowledgements}
We gratefully acknowledge D. Degli-Esposti, D. Michalak and M. Mehmandoost for sharing their expertise on the underlying physics and for their valuable advice. Furthermore, we thank S. L. de Snoo for software support and all the members of the Veldhorst, Vandersypen and Scappucci group for many stimulating discussions.

We acknowledge funding by Intel Corporation. This work is part of the ’Quantum Inspire – the Dutch Quantum Computer in the Cloud’ project (with project number [NWA.1292.19.194]) of the NWA research program ’Research on Routes by Consortia (ORC)’, which is funded by the Netherlands Organization for Scientific Research (NWO).

\section*{Competing interest}
M. Veldhorst is inventor on a patent application related to this work (PCT/N L2022/050377), filling date 30 June 2022. The other authors declare no competing financial interest.

\clearpage
\onecolumngrid

\renewcommand\thefigure{S\arabic{figure}}
\setcounter{figure}{0}
\renewcommand{\thetable}{S\arabic{table}}
\setcounter{table}{0}
\renewcommand{\thesection}{S\arabic{section}}
\setcounter{section}{0}


\begin{center}
\Huge{Supplementary Information}
\end{center}


\section{Stress voltage induced crosstalk}
\label{sup-sec:crosstalk}

\begin{figure}[hbt]
\centering
\includegraphics[]{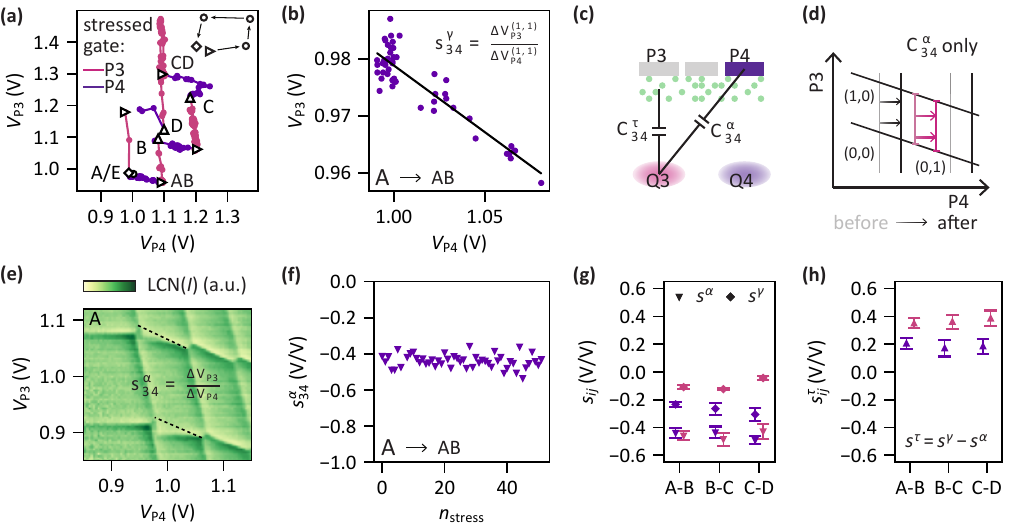}
\caption{\textbf{Stress voltage induced crosstalk on quantum dots.} \textbf{(a)}~Trajectory of the (1,1) charge state center point $\textbf{\textit{V}}^{(1,1)}$ in the $(V_{\rm P3},V_{\rm P4})$ plane during the tuning experiment shown in Fig.~\ref{fig:Fig2} of the main text (identical to Fig.~\ref{fig:Fig2}.c). \textbf{(b)}~Part of the trajectory between the points A and AB. The black line is a linear fit to the data to determine the slope $s^{\gamma}_{34}$ that quantifies the crosstalk of plunger gate P4 on quantum dot Q3. \textbf{(c)} Illustration of a device cross section portraying the capacitive effect of the plunger gate voltage $V_{\rm P4}$ on the potential of quantum dot Q3 ($C^{\alpha}_{34}$) and the crosstalk effect of applying stress voltages to plunger gate P4 on the potential of quantum dot Q3 ($C^{\tau}_{34}$). \textbf{(d)}~Schematic charge stability diagram illustrating how the charge transition voltages of quantum dot Q3 shift when changing the voltage on plunger gate P4. Grey lines indicate the charge transition lines before and black lines after changing the potential of Q4 through applying stress voltages. For illustration purposes the interdot coupling between Q3 and Q4 and the capacitive coupling of P3 onto Q4 are neglected. \textbf{(e)}~Example charge stability diagram taken at point A in \textbf{(a)}. The slope $s^{\alpha}_{34}$ of the transition lines of Q3 (black dashed lines) are determined as a measure for the relative capacitive effect of plunger gate P4 onto the potential of quantum dot Q3. To ensure robustness against distortions from charge latching effects, the Q3 charge transition lines are defined as the lines connecting the respective triple charge degeneracy points. \textbf{(f)}~All extracted $s^{\alpha}_{34}$ during the tuning from point A to AB in \textbf{(a)}. \textbf{(g)}~Crosstalk $s^{\gamma}_{ij}$ caused by stressing plunger gate P$i$ (diamonds) and cross-capacitance effect $s^{\alpha}_{ij}$ of plunger gate voltage $V_{{\rm P}j}$ (downward pointing triangles) onto the potential of quantum dot Q$i$ along the trajectory in \textbf{(a)}. Between C and CD and CD and D only the last ten points are fitted to extract $s^{\gamma}_{ij}$. Due to a limited number of data points no values are shown for the tuning between D and E. \textbf{(h)}~Stress voltage induced crosstalk effect $s^{\tau}_{ij}$ of plunger gate P$j$ onto the potential of quantum dot Q$i$ corrected for the capacitive coupling of plunger gate P$j$ onto the potential of quantum dot Q$i$.}
\label{fig:crosstalk}
\end{figure}

A stress voltage applied to a plunger gate P$j$ not only alters the potential of the quantum dot Q$j$ located directly underneath it but also affects neighbouring quantum dots Q$i$. We investigate this crosstalk by further analyzing the tuning of the Q3-Q4 double quantum dot presented in Fig.~\ref{fig:Fig2} of the main text. Fig.~\ref{sup-sec:crosstalk}.a shows the trajectory of the center $\textbf{\textit{V}}^{(1,1)}$ of the (1,1) charge state region in the $(V_{\rm P3},V_{\rm P4})$ plane (same as Fig.~\ref{fig:Fig2}.c of the main text). The crosstalk manifests as a deviation from perfectly horizontal or vertical progressions of $\textbf{\textit{V}}^{(1,1)}$. We quantify it by applying a linear regression as exemplary shown in Fig.~\ref{sup-sec:crosstalk}.b for the section from A to AB. The extracted slope $s^{\gamma}_{34}$ is a measure for the crosstalk of plunger gate P4 onto quantum dot Q3.

Two mechanisms can explain the observed crosstalk as illustrated in Fig.~\ref{sup-sec:crosstalk}.c: (1) Tuning the potential landscape of Q4 through the application of stress voltages also affects the potential of Q3 even if all gate voltages are reset to their initial value afterwards. For instance, this effect could be caused by the (de)charging of traps at the interface that capacitively couple to Q3 ($C^{\tau}_{34}$)~\cite{Lu2011,Huang2014,Laroche2015,Chou2018,Su2019}. (2) $V^{(1,1)}_{{\rm P}3}$ is defined as the middle point between the (1,0)-(1,1) and (1,1)-(1,2) charge transition at $V_{\rm P4}=V^{(1,1)}_{{\rm P}4}$ (and vice versa). Due to the capacitive coupling of P4 onto Q3 ($C^{\alpha}_{34}$) a shift in $V^{(1,1)}_{{\rm P}4}$ is therefore also reflected in $V^{(1,1)}_{{\rm P}3}$. Fig.~\ref{sup-sec:crosstalk}.d portrays the mechanism. It shows a schematic charge stability diagram before (grey charge transition lines) and after (black charge transition lines) tuning the potential below P4 through the application of stress voltages. As the Q3 charge transition lines are tilted by the cross-capacitance $C^{\alpha}_{34}$, a change in $V^{(1,1)}_{{\rm P}4}$ also results in a change of $V^{(1,1)}_{{\rm P}3}$ (center point of the light and dark pink vertical bar).

To quantify the latter effect we determine the slope $s^{\alpha}_{34}$ of the Q3 charge transition lines at the (1,1) charge region. Fig.~\ref{sup-sec:crosstalk}.e depicts an exemplary charge stability diagram during the tuning process with the respective Q3 charge transition lines indicated by dashed lines. All extracted $s^{\alpha}_{34}$ between the points A and AB in Fig.~\ref{sup-sec:crosstalk}.a are plotted in Fig.\ref{sup-sec:crosstalk}.f. We find that $s^{\alpha}_{34}$ remains constant throughout the entire stress voltage sequence from A to AB.

The same analysis steps are repeated for all sub parts between A and D of the trajectory in Fig.~\ref{sup-sec:crosstalk}.a. Fig.~\ref{sup-sec:crosstalk}.g summarizes all $s^{\gamma}_{ij}$ (diamonds) and $s^{\alpha}_{ij}$ (downward pointing triangles). The magnitude of the cross-capacitance effect $s^{\alpha}_{ij}$ is consistently larger than the magnitude of the measured crosstalk $s^{\gamma}_{ij}$. To estimate the stress voltage crosstalk $s^{\tau}_{ij}$ solely caused by shifts of the intrinsic potential we subtract $s^{\alpha}_{ij}$ from $s^{\gamma}_{ij}$ and plot the difference in Fig.~\ref{sup-sec:crosstalk}.h. We find a positive voltage stress related crosstalk, which has a similar magnitude as the capacitive effect $s^{\alpha}_{ij}$. As $s^{\tau}_{ij}$ and $s^{\alpha}_{ij}$ have a different sign they partially cancel each other and lead to a reduced effective crosstalk $s^{\gamma}_{ij}$ when applying stress voltage sequences.

\section{Additional time traces recorded after applying stress voltages}
\label{sup-sec:time-traces}
\begin{figure}[htb]
\centering
\includegraphics[]{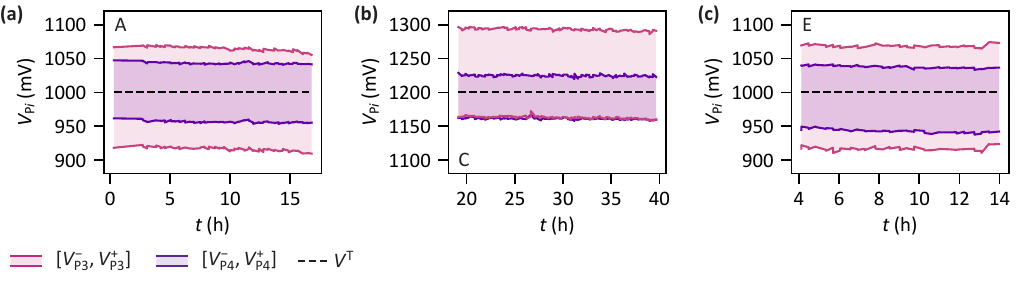}
\caption{\textbf{Additional time traces after applying stress voltage sequences.} \textbf{(a)-(c)} Time traces of the voltage ranges $[V_{{\rm P}i}^{-},V_{{\rm P}i}^{+}]$ after the application of a stress voltage sequence. \textbf{(a)}, \textbf{(b)} and \textbf{(c)} are recorded after tuning to the target points A, C and E  as presented in Fig.~\ref{fig:Fig2} in the main text, respectively. $t$ is the time after the application of the last stress voltage. \textbf{(a)} is identical to Fig.~\ref{fig:Fig3}.a in main text. Note that the underlying charge stability diagram measurements were interleaved with charge noise measurements on the sensor (see supplementary section~\ref{sup-sec:charge-noise}).}
\label{fig:time-traces}
\end{figure}

Fig.~\ref{sup-sec:time-traces} shows two additional time traces not shown in Fig.~\ref{fig:Fig3} in the main text. Note that in Fig.~\ref{sup-sec:time-traces}.b and c the recording of the time traces was started 20~h and 4~h after the application of the last stress voltage, respectively. The additional curves confirm that after the application of a stress voltage tuning the system remains in a (1,1) charge state for 40~h at least only exhibiting small progressive drifts.

\section{Charge noise after applying stress voltages}
\label{sup-sec:charge-noise}

\begin{figure}[htb]
\centering
\includegraphics[]{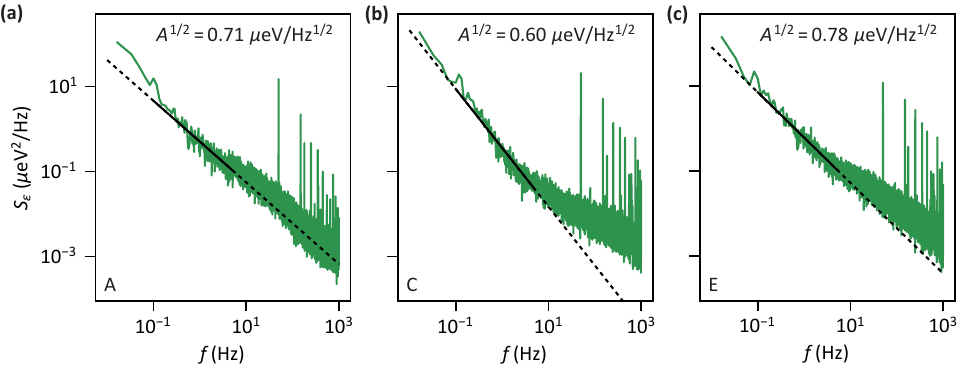}
\caption{\textbf{Sensor charge noise after applying stress voltages.} \textbf{(a)} Power spectral density (PSD) extracted from sensor current time traces recorded after tuning to point A in Fig.~\ref{fig:Fig2}.b of the main text. $S_{\epsilon}=\alpha^2S_{I}/|dI/dV_{\rm S1}|$ with $\alpha$ the lever arm of sensor plunger gate S1 extracted from coulomb diamonds, $|dI/dV_{\rm S1}|$ the maximum slope of the coulomb peak and $S_{I}$ the PSD of the current through the sensor~\cite{Connors2019}. For the measurement the sensor plunger voltage $V_{\rm S1}$ is tuned to the Coulomb peak flank, the voltage for which the sensing quantum dot is most sensitive to potential fluctuations. The black line is a fit to $S_{\epsilon}$ between 0.1~Hz and 5~Hz with $S^{\rm fit}_{\epsilon}=A\times f^{-\kappa}$. The noise amplitude $A$ at 1~Hz is given in the upper right. $\kappa=0.96$ \textbf{(b)} and \textbf{(c)} Same as \textbf{(a)} but recorded after reaching target point C and E in Fig.~\ref{fig:Fig2}.b of the main text, respectively. $\kappa=1.38$ for C and $\kappa=1.07$ for E.}
\label{fig:chrg_noise}
\end{figure}

As the presented tuning procedure might alter the configuration of charge traps in the heterostructure (see supplementary section~\ref{sup-sec:physics}) we investigate the system charge noise after applying stress voltages. Specifically, we measure time traces of the current through the sensing quantum dot (underneath S1) and compute the power spectral density (PSD). To obtain maximum sensitivity of the sensor current to potential fluctuations we tune the sensor plunger gate voltage $V_{\rm S1}$ to the flank of a Coulomb peak. Fig.~\ref{fig:chrg_noise}.a, b and c depict PSD spectra obtained after tuning to the target point A, C and E in Fig.~\ref{fig:Fig2}.b, respectively. Note that target points A and C are reached by applying positively signed stress voltages and target point E is reached by applying negatively signed stress voltages. The charge noise curves follow the typical $1/f$ frequency dependence. Therefore we fit them between 0.1~Hz and 5~Hz with $S^{\rm fit}_{\epsilon}=A\times f^{-\kappa}$ (black line). We find noise amplitudes of $\sqrt{A}=0.71~\mu{\rm eV}/{\rm Hz}^{1/2}$, $\sqrt{A}=0.60~\mu{\rm eV}/{\rm Hz}^{1/2}$ and $\sqrt{A}=0.78~\mu{\rm eV}/{\rm Hz}^{1/2}$ as well as exponents $\kappa=0.96$. $\kappa=1.38$ and $\kappa=1.07$ for target point A, C and E, respectively. These values are comparable to charge noise amplitudes in Si/SiGe reported in the literature~\cite{Connors2019, Struck2020, Connors2022} and charge noise values measured in the same device during an earlier cooldown~\cite{Esposti2023}. Thus, we find no indication that a spin qubit implemented in a stressed quantum dot would be impaired by a degraded noise environment. However, further research is required as the charge noise sensed by the sensor might not be representative for the charge noise affecting qubits that are tuned in the quantum dots.

\section{Identification of the four quantum dots}
\label{sup-sec:four-QD-identification}

\begin{figure}[htb]
\centering
\includegraphics[]{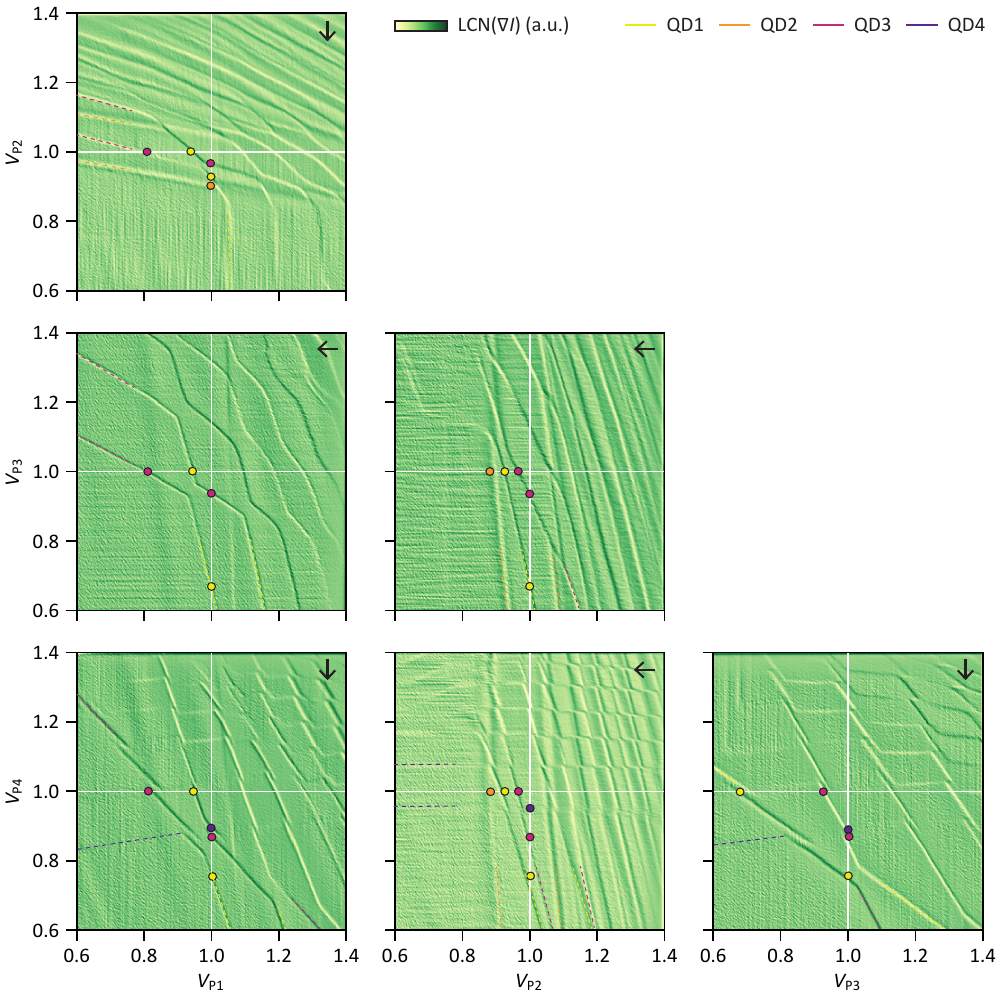}
\caption{\textbf{Charge stability diagrams identifying the quantum dots Q1-Q4.} Charge stability diagrams recorded for all plunger gate combinations. The arrow in the upper left corner indicates the sweep direction. The center point (crossing point of the white lines) for each charge stability diagram corresponds to the same voltage configuration with $V_{\rm P1}=V_{\rm P2}=V_{\rm P3}=V_{\rm P4}=1~$V. Horizontal white lines mark identical line cuts in the gate voltage space inside each column of charge stability diagrams. Vertical white lines mark identical line cuts in the gate voltage space inside each row of charge stability diagrams. Colored dashed lines indicate charge transitions and colored circles mark crossing points of the charge transitions with the white lines. Each color refers to a quantum dot as indicated by the legend in the upper right. To enhance the visibility of the charge transition lines the derivative of the sensor current was taken ad a local contrast normalization (LCN) was applied.}
\label{fig:four-QD-identification}
\end{figure}

In order to identify the quantum dots visible in Fig.~\ref{fig:Fig5} of the main text we measure multiple charge stability diagrams by sweeping all pairwise combinations of the device plunger gate voltages. The obtained charge stability diagrams are plotted in Fig.~\ref{fig:four-QD-identification}. The center left and bottom center panel are identical with the charge stability diagrams shown in Fig.~\ref{fig:Fig5} of the main text. All maps are obtained at the same gate voltage configuration and at their center point all plunger gates are set to 1~V.

The charge stability diagrams can be analyzed starting from one charge transition line, e.g. the first vertical charge transition line in the center left panel (indicated by a yellow dashed line). Due to its strong coupling to plunger gate P1 we identify it as a charge transition line of quantum dot Q1. We mark the crossing point of this Q1 charge transition line with the $V_{\rm P1}=1~$V line (vertical white line) by a yellow circle. Then we place another yellow circle marker at identical $V_{\rm P3}$ on the $V_{\rm P2}=1~$V line in the center panel of the figure. The vertical white lines inside one row of figure panels are identical line cuts in the gate voltage space. Therefore both marked points identify the same charge transition line of the same quantum dot (Q1). Analogously two charge stability diagrams in one column of figure panels can be compared. By repeating the process for all neighbouring charge stability diagrams one can identify the charge transition lines of four quantum dots Q1-Q4. Note that the charge transition lines of quantum dot Q4 (purple) latch when the sweep direction (black arrow in the upper right of each panel) is nearly perpendicular to the charge transition lines. Therefore the crossing point of the first Q4 charge transition line with the $V_{\rm P1}=1~$V line in the bottom left panel and the crossing point with the $V_{\rm P3}=1~$V line in the bottom right panel differ from the crossing point with the $V_{\rm P2}=1~$V line in the bottom center panel. Furthermore, in the left column another nearly vertical charge transition line is visible in the background. However, it shows negligible coupling to the other charge transition lines and likely is a signature of a spurious defect quantum dot outside but close to the active device region.

\section{Underlying physical mechanisms}
\label{sup-sec:physics}
Applying a stress voltage to a selected gate electrode possibly alters the occupation of charge traps in the gate dielectrics and heterostructure directly underneath~\cite{Lu2011,Huang2014,Laroche2015,Chou2018,Su2019}. As the electric field bends the conduction band electrons might tunnel into or out of these charge traps. Removing the stress voltage then effectively freezes their occupation which permanently alters the intrinsic potential landscape. Charge traps can be present in the oxide layer~\cite{Goetzberger1968,Poindexter1988,Lenahan1998,Stesmans2014}, originate from unpassivated silicon and germanium dangling bonds~\cite{Poindexter1988,Lenahan1998,Stesmans2014} or arise from mechanical stress induced by the deposition of metallic gate electrodes~\cite{Thorbeck2015,Stein2021}. Furthermore, also the relocation of mobile ions might change the intrinsic potential~\cite{Vanheusden1998}. Note that these processes in general are independent of the quantum well material itself and stress-voltage-controlled shifts of the intrinsic potential also have been observed in Ge/SiGe heterosturctures~\cite{Su2017,Meyer2023}.

\section{Raw data underlying Fig.~\ref{fig:Fig1}-\ref{fig:Fig4} of the main text}
\label{sup-sec:raw-fig1-4}

\begin{figure}[H]
\centering
\includegraphics[]{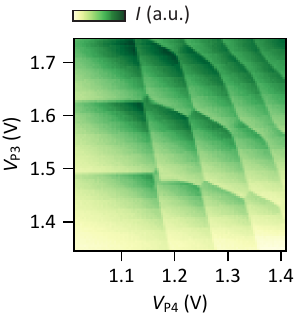}
\caption{\textbf{Raw data underlying Fig.~\ref{fig:Fig1}.b. of the main text}}
\label{fig:fig1b-raw}
\end{figure}

\begin{figure}[H]
\centering
\includegraphics[]{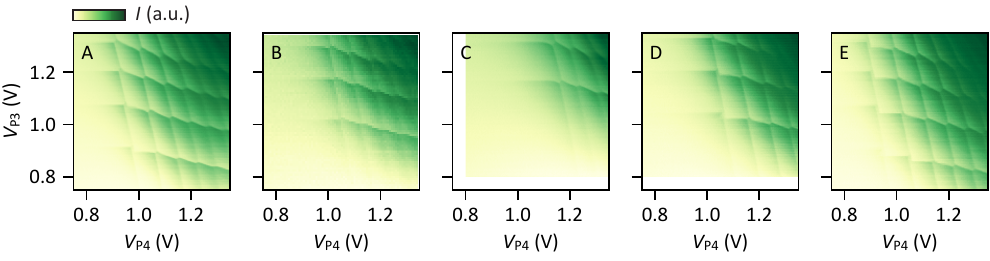}
\caption{\textbf{Raw data underlying Fig.~\ref{fig:Fig3}.f of the main text.}}
\label{fig:fig2f-raw}
\end{figure}

\begin{figure}[H]
\centering
\includegraphics[]{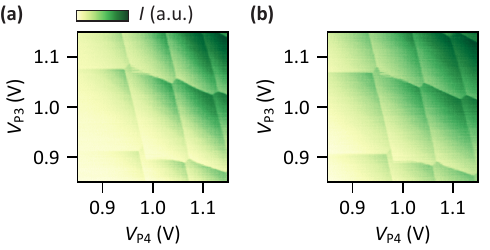}
\caption{\textbf{Raw data underlying Fig.~\ref{fig:Fig3}.b of the main text.} \textbf{(a)}~Charge stability diagram taken at the beginning of the time trace shown in Fig.~\ref{fig:Fig3}.a of the main text. \textbf{(b)}~Charge stability diagram taken at the end of the time trace shown in Fig.~\ref{fig:Fig3}.a of the main text.}
\label{fig:fig3b-raw}
\end{figure}

\begin{figure}[H]
\centering
\includegraphics[]{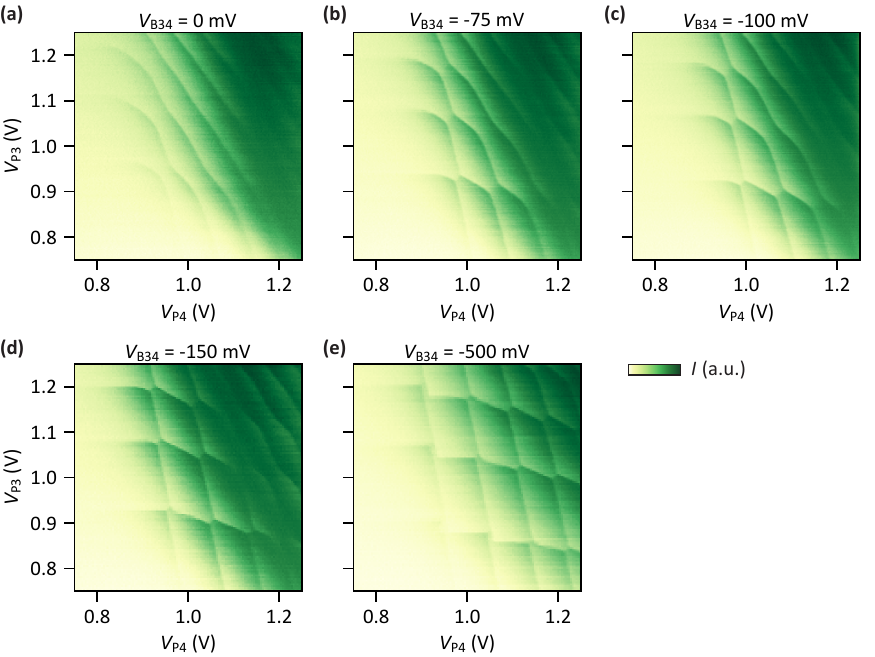}
\caption{\textbf{Raw data underlying Fig.~\ref{fig:Fig4} of the main text.} \textbf{(a)}-\textbf{(e)} correspond to Fig.~\ref{fig:Fig4}.a-e of the main text, respectively.}
\label{fig:fig4-raw}
\end{figure}

Fig.~\ref{fig:fig1b-raw}, \ref{fig:fig2f-raw}, \ref{fig:fig3b-raw}, and \ref{fig:fig4-raw} display the unprocessed charge stability diagram data underlying Fig.~\ref{fig:Fig1}.b, \ref{fig:Fig2}.f, \ref{fig:Fig3}.b, and \ref{fig:Fig4}.a-e of the main text, respectively.

\section{Raw data underlying Fig.~\ref{fig:Fig5} of the main text}
\label{sup-sec:raw-fig5}

\begin{figure}[H]
\centering
\includegraphics[]{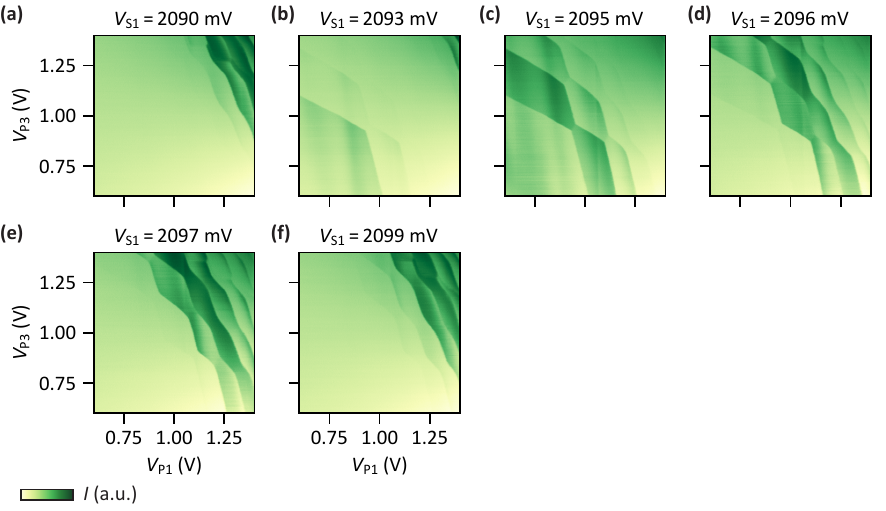}
\caption{\textbf{Charge stability diagrams underlying Fig.~\ref{fig:Fig5}.a.} \textbf{(a)-(f)} Multiple charge stability diagrams showing charge transition lines of quantum dot Q1 and Q3. Maps are taken at various sensor gate S1 voltages as indicated above the plots.}
\label{fig:fig5a-raw}
\end{figure}

\begin{figure}[H]
\centering
\includegraphics[]{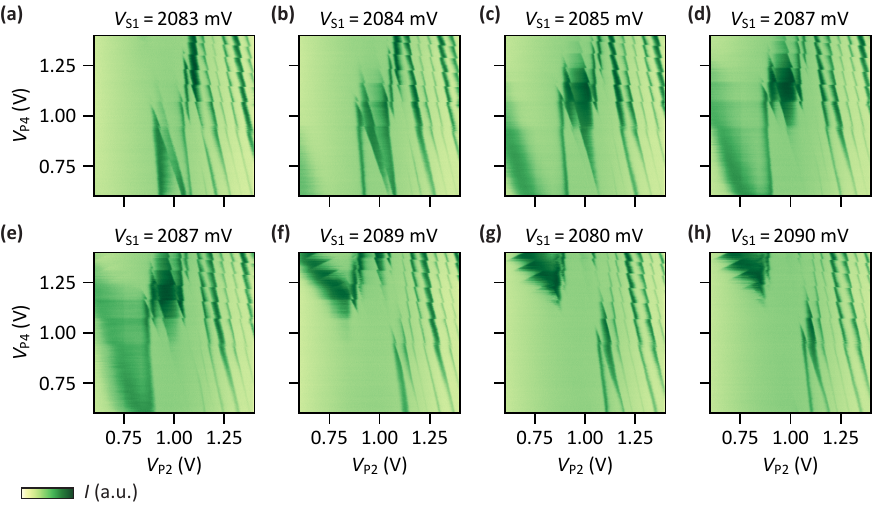}
\caption{\textbf{Charge stability diagrams underlying Fig.~\ref{fig:Fig5}.b.} \textbf{(a)-(e)} Multiple charge stability diagrams showing charge transition lines of quantum dot Q1-4. Maps are taken at various sensor gate S1 voltages as indicated above the plots.}
\label{fig:fig5b-raw}
\end{figure}

Fig.~\ref{fig:fig5a-raw} and Fig.~\ref{fig:fig5b-raw} show the unprocessed charge stability diagram data underlying Fig.~\ref{fig:Fig5}.a and b of the main text, respectively. Each map is recorded at a different sensor gate S1 voltage to account for the cross-capacitance effect of the plunger gates on the sensing dot potential which limits the sensing dot sensitivity to small plunger gate voltage ranges.

\begin{figure}[H]
\centering
\includegraphics[]{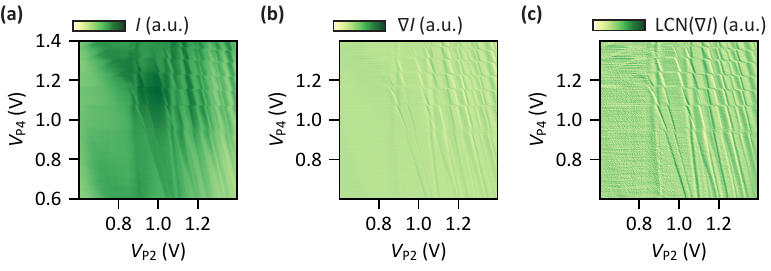}
\caption{\textbf{Processing of the data underlying Fig.~\ref{fig:Fig5}.b.} \textbf{(a)} Sum of the sensor response $I$ of the charge stability diagrams shown in Fig.~\ref{fig:fig5b-raw}. \textbf{(b)} Gradient $\nabla I$ of the data shown in \textbf{(a)}. \textbf{(c)} Final signal ${\rm LCN}(\nabla I)$ after applying a local contrast normalization to the map shown in \textbf{(b)}.}
\label{fig:fig5b-process}
\end{figure}

We combine the charge stability diagrams by summing up the sensor current signals as exemplary shown in Fig.~\ref{fig:fig5b-process}.a for the data shown in Fig.~\ref{fig:fig5b-raw}. Afterwards, the signal gradient $\nabla I$ is calculated as depicted in Fig.~\ref{fig:fig5b-process}.b. Finally, a local contrast normalization (see methods section) is applied to allow for an eased identification of charge transition lines across the full map. Fig.~\ref{fig:fig5b-process}.c depicts the resulting charge stability diagram which is identical to the charge stability diagram shown in Fig.~\ref{fig:Fig5}.b of the main text.

\section{Overview of applied gate voltage configurations}
\label{sup-sec:voltage-configs}

\begin{figure}[H]
\centering
\includegraphics[]{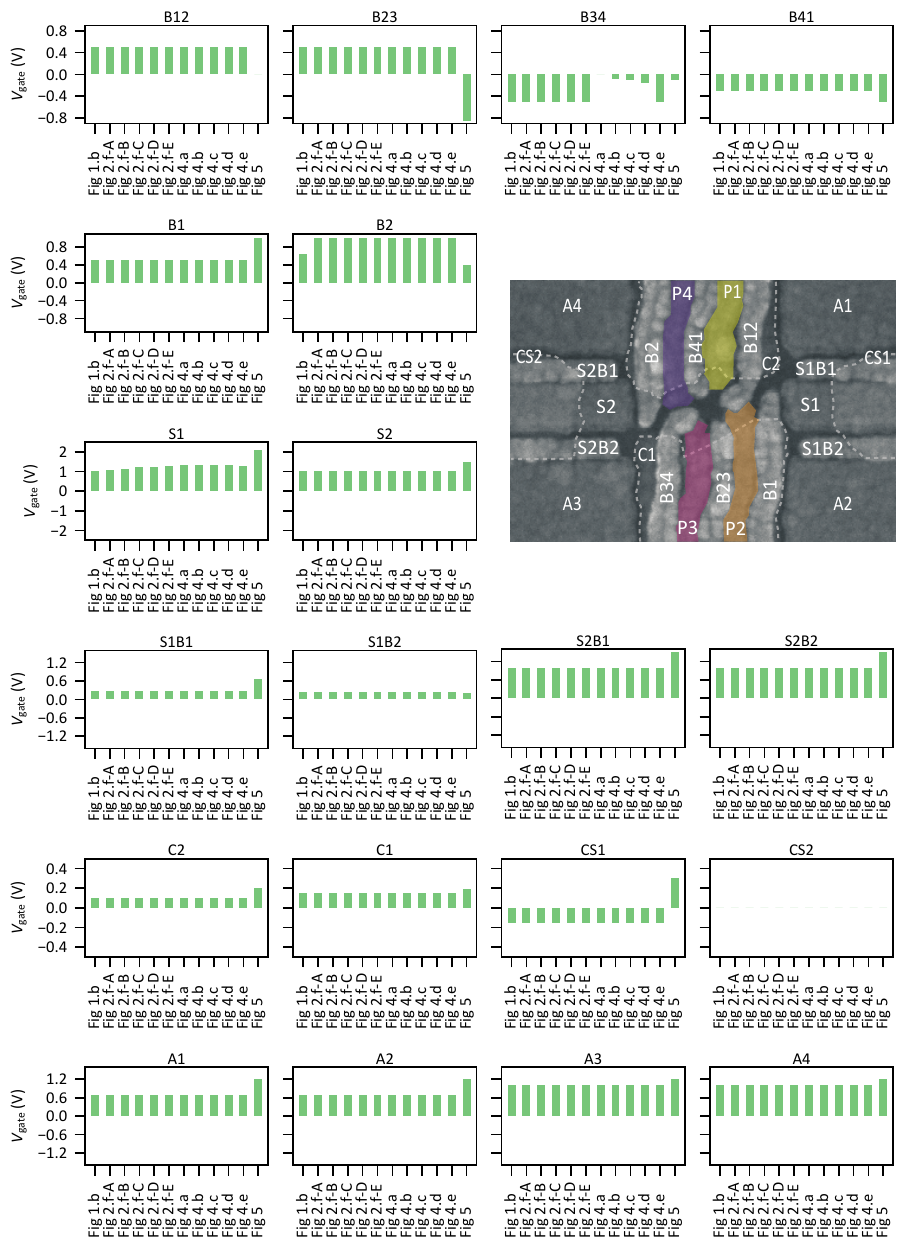}
\caption{\textbf{Gate voltage evolution during the presented experiments.} Each panel shows the gate voltage evolution of a single gate during the experiments presented in the figures of the main text as given on the x-axis. Note that $V_{\rm S2C}=0~$V during all experiments. The inset shows an SEM image of a device nominally identical to the one under study. Confinement gates are outlined by a white dashed line. Labels indicate the gate electrode naming convention utilized throughout the manuscript and in the panels of this figure.}
\label{fig:voltage-configs}
\end{figure}



\bibliography{lib_manuscript}

\begin{thebibliography}{57}%
\makeatletter
\providecommand \@ifxundefined [1]{%
 \@ifx{#1\undefined}
}%
\providecommand \@ifnum [1]{%
 \ifnum #1\expandafter \@firstoftwo
 \else \expandafter \@secondoftwo
 \fi
}%
\providecommand \@ifx [1]{%
 \ifx #1\expandafter \@firstoftwo
 \else \expandafter \@secondoftwo
 \fi
}%
\providecommand \natexlab [1]{#1}%
\providecommand \enquote  [1]{``#1''}%
\providecommand \bibnamefont  [1]{#1}%
\providecommand \bibfnamefont [1]{#1}%
\providecommand \citenamefont [1]{#1}%
\providecommand \href@noop [0]{\@secondoftwo}%
\providecommand \href [0]{\begingroup \@sanitize@url \@href}%
\providecommand \@href[1]{\@@startlink{#1}\@@href}%
\providecommand \@@href[1]{\endgroup#1\@@endlink}%
\providecommand \@sanitize@url [0]{\catcode `\\12\catcode `\$12\catcode
  `\&12\catcode `\#12\catcode `\^12\catcode `\_12\catcode `\%12\relax}%
\providecommand \@@startlink[1]{}%
\providecommand \@@endlink[0]{}%
\providecommand \url  [0]{\begingroup\@sanitize@url \@url }%
\providecommand \@url [1]{\endgroup\@href {#1}{\urlprefix }}%
\providecommand \urlprefix  [0]{URL }%
\providecommand \Eprint [0]{\href }%
\providecommand \doibase [0]{https://doi.org/}%
\providecommand \selectlanguage [0]{\@gobble}%
\providecommand \bibinfo  [0]{\@secondoftwo}%
\providecommand \bibfield  [0]{\@secondoftwo}%
\providecommand \translation [1]{[#1]}%
\providecommand \BibitemOpen [0]{}%
\providecommand \bibitemStop [0]{}%
\providecommand \bibitemNoStop [0]{.\EOS\space}%
\providecommand \EOS [0]{\spacefactor3000\relax}%
\providecommand \BibitemShut  [1]{\csname bibitem#1\endcsname}%
\let\auto@bib@innerbib\@empty
\bibitem [{\citenamefont {Lawrie}\ \emph {et~al.}(2023)\citenamefont {Lawrie},
  \citenamefont {Rimbach-Russ}, \citenamefont {Riggelen}, \citenamefont
  {Hendrickx}, \citenamefont {Snoo}, \citenamefont {Sammak}, \citenamefont
  {Scappucci}, \citenamefont {Helsen},\ and\ \citenamefont
  {Veldhorst}}]{Lawrie2023}%
  \BibitemOpen
  \bibfield  {author} {\bibinfo {author} {\bibfnamefont {W.~I.~L.}\
  \bibnamefont {Lawrie}}, \bibinfo {author} {\bibfnamefont {M.}~\bibnamefont
  {Rimbach-Russ}}, \bibinfo {author} {\bibfnamefont {F.~v.}\ \bibnamefont
  {Riggelen}}, \bibinfo {author} {\bibfnamefont {N.~W.}\ \bibnamefont
  {Hendrickx}}, \bibinfo {author} {\bibfnamefont {S.~L.~d.}\ \bibnamefont
  {Snoo}}, \bibinfo {author} {\bibfnamefont {A.}~\bibnamefont {Sammak}},
  \bibinfo {author} {\bibfnamefont {G.}~\bibnamefont {Scappucci}}, \bibinfo
  {author} {\bibfnamefont {J.}~\bibnamefont {Helsen}},\ and\ \bibinfo {author}
  {\bibfnamefont {M.}~\bibnamefont {Veldhorst}},\ }\bibfield  {title} {\bibinfo
  {title} {Simultaneous single-qubit driving of semiconductor spin qubits at
  the fault-tolerant threshold},\ }\href@noop {} {\bibfield  {journal}
  {\bibinfo  {journal} {Nature Communications}\ }\textbf {\bibinfo {volume}
  {14}},\ \bibinfo {pages} {3617} (\bibinfo {year} {2023})}\BibitemShut
  {NoStop}%
\bibitem [{\citenamefont {M{\k{a}}dzik}\ \emph {et~al.}(2022)\citenamefont
  {M{\k{a}}dzik}, \citenamefont {Asaad}, \citenamefont {Youssry}, \citenamefont
  {Joecker}, \citenamefont {Rudinger}, \citenamefont {Nielsen}, \citenamefont
  {Young}, \citenamefont {Proctor}, \citenamefont {Baczewski}, \citenamefont
  {Laucht}, \citenamefont {Schmitt}, \citenamefont {Hudson}, \citenamefont
  {Itoh}, \citenamefont {Jakob}, \citenamefont {Johnson}, \citenamefont
  {Jamieson}, \citenamefont {Dzurak}, \citenamefont {Ferrie}, \citenamefont
  {Blume-Kohout},\ and\ \citenamefont {Morello}}]{Madzik2022}%
  \BibitemOpen
  \bibfield  {author} {\bibinfo {author} {\bibfnamefont {M.}~\bibnamefont
  {M{\k{a}}dzik}}, \bibinfo {author} {\bibfnamefont {S.}~\bibnamefont {Asaad}},
  \bibinfo {author} {\bibfnamefont {A.}~\bibnamefont {Youssry}}, \bibinfo
  {author} {\bibfnamefont {B.}~\bibnamefont {Joecker}}, \bibinfo {author}
  {\bibfnamefont {K.~M.}\ \bibnamefont {Rudinger}}, \bibinfo {author}
  {\bibfnamefont {E.}~\bibnamefont {Nielsen}}, \bibinfo {author} {\bibfnamefont
  {K.~C.}\ \bibnamefont {Young}}, \bibinfo {author} {\bibfnamefont {T.~J.}\
  \bibnamefont {Proctor}}, \bibinfo {author} {\bibfnamefont {A.~D.}\
  \bibnamefont {Baczewski}}, \bibinfo {author} {\bibfnamefont {A.}~\bibnamefont
  {Laucht}}, \bibinfo {author} {\bibfnamefont {V.}~\bibnamefont {Schmitt}},
  \bibinfo {author} {\bibfnamefont {F.~E.}\ \bibnamefont {Hudson}}, \bibinfo
  {author} {\bibfnamefont {K.~M.}\ \bibnamefont {Itoh}}, \bibinfo {author}
  {\bibfnamefont {A.~M.}\ \bibnamefont {Jakob}}, \bibinfo {author}
  {\bibfnamefont {B.~C.}\ \bibnamefont {Johnson}}, \bibinfo {author}
  {\bibfnamefont {D.~N.}\ \bibnamefont {Jamieson}}, \bibinfo {author}
  {\bibfnamefont {A.~S.}\ \bibnamefont {Dzurak}}, \bibinfo {author}
  {\bibfnamefont {C.}~\bibnamefont {Ferrie}}, \bibinfo {author} {\bibfnamefont
  {R.}~\bibnamefont {Blume-Kohout}},\ and\ \bibinfo {author} {\bibfnamefont
  {A.}~\bibnamefont {Morello}},\ }\bibfield  {title} {\bibinfo {title}
  {Precision tomography of a three-qubit donor quantum processor in silicon},\
  }\href@noop {} {\bibfield  {journal} {\bibinfo  {journal} {Nature}\ }\textbf
  {\bibinfo {volume} {601}},\ \bibinfo {pages} {348} (\bibinfo {year}
  {2022})}\BibitemShut {NoStop}%
\bibitem [{\citenamefont {Noiri}\ \emph
  {et~al.}(2022{\natexlab{a}})\citenamefont {Noiri}, \citenamefont {Takeda},
  \citenamefont {Nakajima}, \citenamefont {Kobayashi}, \citenamefont {Sammak},
  \citenamefont {Scappucci},\ and\ \citenamefont {Tarucha}}]{Noiri2022}%
  \BibitemOpen
  \bibfield  {author} {\bibinfo {author} {\bibfnamefont {A.}~\bibnamefont
  {Noiri}}, \bibinfo {author} {\bibfnamefont {K.}~\bibnamefont {Takeda}},
  \bibinfo {author} {\bibfnamefont {T.}~\bibnamefont {Nakajima}}, \bibinfo
  {author} {\bibfnamefont {T.}~\bibnamefont {Kobayashi}}, \bibinfo {author}
  {\bibfnamefont {A.}~\bibnamefont {Sammak}}, \bibinfo {author} {\bibfnamefont
  {G.}~\bibnamefont {Scappucci}},\ and\ \bibinfo {author} {\bibfnamefont
  {S.}~\bibnamefont {Tarucha}},\ }\bibfield  {title} {\bibinfo {title} {Fast
  universal quantum gate above the fault-tolerance threshold in silicon},\
  }\href@noop {} {\bibfield  {journal} {\bibinfo  {journal} {Nature}\ }\textbf
  {\bibinfo {volume} {601}},\ \bibinfo {pages} {338} (\bibinfo {year}
  {2022}{\natexlab{a}})}\BibitemShut {NoStop}%
\bibitem [{\citenamefont {Xue}\ \emph {et~al.}(2022)\citenamefont {Xue},
  \citenamefont {Russ}, \citenamefont {Samkharadze}, \citenamefont {Undseth},
  \citenamefont {Sammak}, \citenamefont {Scappucci},\ and\ \citenamefont
  {Vandersypen}}]{Xue2022}%
  \BibitemOpen
  \bibfield  {author} {\bibinfo {author} {\bibfnamefont {X.}~\bibnamefont
  {Xue}}, \bibinfo {author} {\bibfnamefont {M.}~\bibnamefont {Russ}}, \bibinfo
  {author} {\bibfnamefont {N.}~\bibnamefont {Samkharadze}}, \bibinfo {author}
  {\bibfnamefont {B.}~\bibnamefont {Undseth}}, \bibinfo {author} {\bibfnamefont
  {A.}~\bibnamefont {Sammak}}, \bibinfo {author} {\bibfnamefont
  {G.}~\bibnamefont {Scappucci}},\ and\ \bibinfo {author} {\bibfnamefont
  {L.~M.~K.}\ \bibnamefont {Vandersypen}},\ }\bibfield  {title} {\bibinfo
  {title} {Quantum logic with spin qubits crossing the surface code
  threshold},\ }\href@noop {} {\bibfield  {journal} {\bibinfo  {journal}
  {Nature}\ }\textbf {\bibinfo {volume} {601}},\ \bibinfo {pages} {343}
  (\bibinfo {year} {2022})}\BibitemShut {NoStop}%
\bibitem [{\citenamefont {Mills}\ \emph {et~al.}(2022)\citenamefont {Mills},
  \citenamefont {Guinn}, \citenamefont {Gullans}, \citenamefont {Sigillito},
  \citenamefont {Feldman}, \citenamefont {Nielsen},\ and\ \citenamefont
  {Petta}}]{Mills2022}%
  \BibitemOpen
  \bibfield  {author} {\bibinfo {author} {\bibfnamefont {A.~R.}\ \bibnamefont
  {Mills}}, \bibinfo {author} {\bibfnamefont {C.~R.}\ \bibnamefont {Guinn}},
  \bibinfo {author} {\bibfnamefont {M.~J.}\ \bibnamefont {Gullans}}, \bibinfo
  {author} {\bibfnamefont {A.}~\bibnamefont {Sigillito}}, \bibinfo {author}
  {\bibfnamefont {M.~M.}\ \bibnamefont {Feldman}}, \bibinfo {author}
  {\bibfnamefont {E.}~\bibnamefont {Nielsen}},\ and\ \bibinfo {author}
  {\bibfnamefont {J.~R.}\ \bibnamefont {Petta}},\ }\bibfield  {title} {\bibinfo
  {title} {Two-qubit silicon quantum processor with operation fidelity
  exceeding 99\%},\ }\href@noop {} {\bibfield  {journal} {\bibinfo  {journal}
  {Science Advances}\ }\textbf {\bibinfo {volume} {8}},\ \bibinfo {pages}
  {eabn5130} (\bibinfo {year} {2022})}\BibitemShut {NoStop}%
\bibitem [{\citenamefont {Petit}\ \emph {et~al.}(2020)\citenamefont {Petit},
  \citenamefont {Eenink}, \citenamefont {Russ}, \citenamefont {Lawrie},
  \citenamefont {Hendrickx}, \citenamefont {Philips}, \citenamefont {Clarke},
  \citenamefont {Vandersypen},\ and\ \citenamefont {Veldhorst}}]{Petit2020}%
  \BibitemOpen
  \bibfield  {author} {\bibinfo {author} {\bibfnamefont {L.}~\bibnamefont
  {Petit}}, \bibinfo {author} {\bibfnamefont {H.~G.~J.}\ \bibnamefont
  {Eenink}}, \bibinfo {author} {\bibfnamefont {M.}~\bibnamefont {Russ}},
  \bibinfo {author} {\bibfnamefont {W.~I.~L.}\ \bibnamefont {Lawrie}}, \bibinfo
  {author} {\bibfnamefont {N.~W.}\ \bibnamefont {Hendrickx}}, \bibinfo {author}
  {\bibfnamefont {S.~G.~J.}\ \bibnamefont {Philips}}, \bibinfo {author}
  {\bibfnamefont {J.~S.}\ \bibnamefont {Clarke}}, \bibinfo {author}
  {\bibfnamefont {L.~M.~K.}\ \bibnamefont {Vandersypen}},\ and\ \bibinfo
  {author} {\bibfnamefont {M.}~\bibnamefont {Veldhorst}},\ }\bibfield  {title}
  {\bibinfo {title} {Universal quantum logic in hot silicon qubits},\
  }\href@noop {} {\bibfield  {journal} {\bibinfo  {journal} {Nature}\ }\textbf
  {\bibinfo {volume} {580}},\ \bibinfo {pages} {355} (\bibinfo {year}
  {2020})}\BibitemShut {NoStop}%
\bibitem [{\citenamefont {Yang}\ \emph {et~al.}(2020)\citenamefont {Yang},
  \citenamefont {Leon}, \citenamefont {Hwang}, \citenamefont {Saraiva},
  \citenamefont {Tanttu}, \citenamefont {Huang}, \citenamefont
  {Camirand~Lemyre}, \citenamefont {Chan}, \citenamefont {Tan}, \citenamefont
  {Hudson}, \citenamefont {Itoh}, \citenamefont {Morello}, \citenamefont
  {Pioro-Ladrière}, \citenamefont {Laucht},\ and\ \citenamefont
  {Dzurak}}]{Yang2020}%
  \BibitemOpen
  \bibfield  {author} {\bibinfo {author} {\bibfnamefont {C.~H.}\ \bibnamefont
  {Yang}}, \bibinfo {author} {\bibfnamefont {R.~C.~C.}\ \bibnamefont {Leon}},
  \bibinfo {author} {\bibfnamefont {J.~C.~C.}\ \bibnamefont {Hwang}}, \bibinfo
  {author} {\bibfnamefont {A.}~\bibnamefont {Saraiva}}, \bibinfo {author}
  {\bibfnamefont {T.}~\bibnamefont {Tanttu}}, \bibinfo {author} {\bibfnamefont
  {W.}~\bibnamefont {Huang}}, \bibinfo {author} {\bibfnamefont
  {J.}~\bibnamefont {Camirand~Lemyre}}, \bibinfo {author} {\bibfnamefont
  {K.~W.}\ \bibnamefont {Chan}}, \bibinfo {author} {\bibfnamefont {K.~Y.}\
  \bibnamefont {Tan}}, \bibinfo {author} {\bibfnamefont {F.~E.}\ \bibnamefont
  {Hudson}}, \bibinfo {author} {\bibfnamefont {K.~M.}\ \bibnamefont {Itoh}},
  \bibinfo {author} {\bibfnamefont {A.}~\bibnamefont {Morello}}, \bibinfo
  {author} {\bibfnamefont {M.}~\bibnamefont {Pioro-Ladrière}}, \bibinfo
  {author} {\bibfnamefont {A.}~\bibnamefont {Laucht}},\ and\ \bibinfo {author}
  {\bibfnamefont {A.~S.}\ \bibnamefont {Dzurak}},\ }\bibfield  {title}
  {\bibinfo {title} {Operation of a silicon quantum processor unit cell above
  one kelvin},\ }\href@noop {} {\bibfield  {journal} {\bibinfo  {journal}
  {Nature}\ }\textbf {\bibinfo {volume} {580}},\ \bibinfo {pages} {350}
  (\bibinfo {year} {2020})}\BibitemShut {NoStop}%
\bibitem [{\citenamefont {Camenzind}\ \emph {et~al.}(2022)\citenamefont
  {Camenzind}, \citenamefont {Geyer}, \citenamefont {Fuhrer}, \citenamefont
  {Warburton}, \citenamefont {Zumbühl},\ and\ \citenamefont
  {Kuhlmann}}]{Camenzind2022}%
  \BibitemOpen
  \bibfield  {author} {\bibinfo {author} {\bibfnamefont {L.~C.}\ \bibnamefont
  {Camenzind}}, \bibinfo {author} {\bibfnamefont {S.}~\bibnamefont {Geyer}},
  \bibinfo {author} {\bibfnamefont {A.}~\bibnamefont {Fuhrer}}, \bibinfo
  {author} {\bibfnamefont {R.~J.}\ \bibnamefont {Warburton}}, \bibinfo {author}
  {\bibfnamefont {D.~M.}\ \bibnamefont {Zumbühl}},\ and\ \bibinfo {author}
  {\bibfnamefont {A.}~\bibnamefont {Kuhlmann}},\ }\bibfield  {title} {\bibinfo
  {title} {A hole spin qubit in a fin field-effect transistor above 4 kelvin},\
  }\href@noop {} {\bibfield  {journal} {\bibinfo  {journal} {Nature
  Electronics}\ }\textbf {\bibinfo {volume} {5}},\ \bibinfo {pages} {178}
  (\bibinfo {year} {2022})}\BibitemShut {NoStop}%
\bibitem [{\citenamefont {Bourdet}\ \emph {et~al.}(2018)\citenamefont
  {Bourdet}, \citenamefont {Hutin}, \citenamefont {Bertrand}, \citenamefont
  {Corna}, \citenamefont {Bohuslavskyi}, \citenamefont {Amisse}, \citenamefont
  {Crippa}, \citenamefont {Maurand}, \citenamefont {Barraud}, \citenamefont
  {Urdampilleta}, \citenamefont {Bäuerle}, \citenamefont {Meunier},
  \citenamefont {Sanquer}, \citenamefont {Jehl}, \citenamefont {Franceschi},
  \citenamefont {Niquet},\ and\ \citenamefont {Vinet}}]{Bourdet2018}%
  \BibitemOpen
  \bibfield  {author} {\bibinfo {author} {\bibfnamefont {L.}~\bibnamefont
  {Bourdet}}, \bibinfo {author} {\bibfnamefont {L.}~\bibnamefont {Hutin}},
  \bibinfo {author} {\bibfnamefont {B.}~\bibnamefont {Bertrand}}, \bibinfo
  {author} {\bibfnamefont {A.}~\bibnamefont {Corna}}, \bibinfo {author}
  {\bibfnamefont {H.}~\bibnamefont {Bohuslavskyi}}, \bibinfo {author}
  {\bibfnamefont {A.}~\bibnamefont {Amisse}}, \bibinfo {author} {\bibfnamefont
  {A.}~\bibnamefont {Crippa}}, \bibinfo {author} {\bibfnamefont
  {R.}~\bibnamefont {Maurand}}, \bibinfo {author} {\bibfnamefont
  {S.}~\bibnamefont {Barraud}}, \bibinfo {author} {\bibfnamefont
  {M.}~\bibnamefont {Urdampilleta}}, \bibinfo {author} {\bibfnamefont
  {C.}~\bibnamefont {Bäuerle}}, \bibinfo {author} {\bibfnamefont
  {T.}~\bibnamefont {Meunier}}, \bibinfo {author} {\bibfnamefont
  {M.}~\bibnamefont {Sanquer}}, \bibinfo {author} {\bibfnamefont
  {X.}~\bibnamefont {Jehl}}, \bibinfo {author} {\bibfnamefont {S.~D.}\
  \bibnamefont {Franceschi}}, \bibinfo {author} {\bibfnamefont {Y.~M.}\
  \bibnamefont {Niquet}},\ and\ \bibinfo {author} {\bibfnamefont
  {M.}~\bibnamefont {Vinet}},\ }\bibfield  {title} {\bibinfo {title}
  {All-electrical control of a hybrid electron spin/valley quantum bit in soi
  cmos technology},\ }\href@noop {} {\bibfield  {journal} {\bibinfo  {journal}
  {IEEE Transactions on Electron Devices}\ }\textbf {\bibinfo {volume} {65}},\
  \bibinfo {pages} {5151} (\bibinfo {year} {2018})}\BibitemShut {NoStop}%
\bibitem [{\citenamefont {Ansaloni}\ \emph {et~al.}(2020)\citenamefont
  {Ansaloni}, \citenamefont {Chatterjee}, \citenamefont {Bohuslavskyi},
  \citenamefont {Bertrand}, \citenamefont {Hutin}, \citenamefont {Vinet},\ and\
  \citenamefont {Kuemmeth}}]{Ansaloni2020}%
  \BibitemOpen
  \bibfield  {author} {\bibinfo {author} {\bibfnamefont {F.}~\bibnamefont
  {Ansaloni}}, \bibinfo {author} {\bibfnamefont {A.}~\bibnamefont
  {Chatterjee}}, \bibinfo {author} {\bibfnamefont {H.}~\bibnamefont
  {Bohuslavskyi}}, \bibinfo {author} {\bibfnamefont {B.}~\bibnamefont
  {Bertrand}}, \bibinfo {author} {\bibfnamefont {L.}~\bibnamefont {Hutin}},
  \bibinfo {author} {\bibfnamefont {M.}~\bibnamefont {Vinet}},\ and\ \bibinfo
  {author} {\bibfnamefont {F.}~\bibnamefont {Kuemmeth}},\ }\bibfield  {title}
  {\bibinfo {title} {Single-electron operations in a foundry-fabricated array
  of quantum dots},\ }\href@noop {} {\bibfield  {journal} {\bibinfo  {journal}
  {Nature Communications}\ }\textbf {\bibinfo {volume} {11}},\ \bibinfo {pages}
  {6399} (\bibinfo {year} {2020})}\BibitemShut {NoStop}%
\bibitem [{\citenamefont {Zwerver}\ \emph {et~al.}(2022)\citenamefont
  {Zwerver}, \citenamefont {Kr\"ahenmann}, \citenamefont {Watson},
  \citenamefont {Lampert}, \citenamefont {George}, \citenamefont
  {Pillarisetty}, \citenamefont {Bojarski}, \citenamefont {Amin}, \citenamefont
  {Amitonov}, \citenamefont {Boter}, \citenamefont {Caudillo}, \citenamefont
  {Corras-Serrano}, \citenamefont {Dehollain}, \citenamefont {Droulers},
  \citenamefont {Henry}, \citenamefont {Kotlyar}, \citenamefont {Lodari},
  \citenamefont {L\"uthi}, \citenamefont {Michalak}, \citenamefont {Mueller},
  \citenamefont {Neyens}, \citenamefont {Roberts}, \citenamefont {Samkharadze},
  \citenamefont {Zheng}, \citenamefont {Zietz}, \citenamefont {Scappucci},
  \citenamefont {Veldhorst}, \citenamefont {Vandersypen},\ and\ \citenamefont
  {Clarke}}]{Zwerver2022}%
  \BibitemOpen
  \bibfield  {author} {\bibinfo {author} {\bibfnamefont {A.-M.~J.}\
  \bibnamefont {Zwerver}}, \bibinfo {author} {\bibfnamefont {T.}~\bibnamefont
  {Kr\"ahenmann}}, \bibinfo {author} {\bibfnamefont {T.~F.}\ \bibnamefont
  {Watson}}, \bibinfo {author} {\bibfnamefont {L.}~\bibnamefont {Lampert}},
  \bibinfo {author} {\bibfnamefont {H.~C.}\ \bibnamefont {George}}, \bibinfo
  {author} {\bibfnamefont {R.}~\bibnamefont {Pillarisetty}}, \bibinfo {author}
  {\bibfnamefont {S.~A.}\ \bibnamefont {Bojarski}}, \bibinfo {author}
  {\bibfnamefont {P.}~\bibnamefont {Amin}}, \bibinfo {author} {\bibfnamefont
  {S.~V.}\ \bibnamefont {Amitonov}}, \bibinfo {author} {\bibfnamefont {J.~M.}\
  \bibnamefont {Boter}}, \bibinfo {author} {\bibfnamefont {R.}~\bibnamefont
  {Caudillo}}, \bibinfo {author} {\bibfnamefont {D.}~\bibnamefont
  {Corras-Serrano}}, \bibinfo {author} {\bibfnamefont {J.~P.}\ \bibnamefont
  {Dehollain}}, \bibinfo {author} {\bibfnamefont {G.}~\bibnamefont {Droulers}},
  \bibinfo {author} {\bibfnamefont {E.~M.}\ \bibnamefont {Henry}}, \bibinfo
  {author} {\bibfnamefont {R.}~\bibnamefont {Kotlyar}}, \bibinfo {author}
  {\bibfnamefont {M.}~\bibnamefont {Lodari}}, \bibinfo {author} {\bibfnamefont
  {F.}~\bibnamefont {L\"uthi}}, \bibinfo {author} {\bibfnamefont {D.~J.}\
  \bibnamefont {Michalak}}, \bibinfo {author} {\bibfnamefont {B.~K.}\
  \bibnamefont {Mueller}}, \bibinfo {author} {\bibfnamefont {S.}~\bibnamefont
  {Neyens}}, \bibinfo {author} {\bibfnamefont {J.}~\bibnamefont {Roberts}},
  \bibinfo {author} {\bibfnamefont {N.}~\bibnamefont {Samkharadze}}, \bibinfo
  {author} {\bibfnamefont {G.}~\bibnamefont {Zheng}}, \bibinfo {author}
  {\bibfnamefont {O.~K.}\ \bibnamefont {Zietz}}, \bibinfo {author}
  {\bibfnamefont {G.}~\bibnamefont {Scappucci}}, \bibinfo {author}
  {\bibfnamefont {M.}~\bibnamefont {Veldhorst}}, \bibinfo {author}
  {\bibfnamefont {L.~M.~K.}\ \bibnamefont {Vandersypen}},\ and\ \bibinfo
  {author} {\bibfnamefont {J.~S.}\ \bibnamefont {Clarke}},\ }\bibfield  {title}
  {\bibinfo {title} {Qubits made by advanced semiconductor manufacturing},\
  }\href@noop {} {\bibfield  {journal} {\bibinfo  {journal} {Nature
  Electronics}\ }\textbf {\bibinfo {volume} {5}},\ \bibinfo {pages} {184}
  (\bibinfo {year} {2022})}\BibitemShut {NoStop}%
\bibitem [{\citenamefont {Hendrickx}\ \emph {et~al.}(2021)\citenamefont
  {Hendrickx}, \citenamefont {Lawrie}, \citenamefont {Russ}, \citenamefont {van
  Riggelen}, \citenamefont {de~Snoo}, \citenamefont {Schouten}, \citenamefont
  {Sammak}, \citenamefont {Scappucci},\ and\ \citenamefont
  {Veldhorst}}]{Hendrickx2021}%
  \BibitemOpen
  \bibfield  {author} {\bibinfo {author} {\bibfnamefont {N.~W.}\ \bibnamefont
  {Hendrickx}}, \bibinfo {author} {\bibfnamefont {W.~I.~L.}\ \bibnamefont
  {Lawrie}}, \bibinfo {author} {\bibfnamefont {M.}~\bibnamefont {Russ}},
  \bibinfo {author} {\bibfnamefont {F.}~\bibnamefont {van Riggelen}}, \bibinfo
  {author} {\bibfnamefont {S.~L.}\ \bibnamefont {de~Snoo}}, \bibinfo {author}
  {\bibfnamefont {R.~N.}\ \bibnamefont {Schouten}}, \bibinfo {author}
  {\bibfnamefont {A.}~\bibnamefont {Sammak}}, \bibinfo {author} {\bibfnamefont
  {G.}~\bibnamefont {Scappucci}},\ and\ \bibinfo {author} {\bibfnamefont
  {M.}~\bibnamefont {Veldhorst}},\ }\bibfield  {title} {\bibinfo {title} {A
  four-qubit germanium quantum processor},\ }\href@noop {} {\bibfield
  {journal} {\bibinfo  {journal} {Nature}\ }\textbf {\bibinfo {volume} {591}},\
  \bibinfo {pages} {580} (\bibinfo {year} {2021})}\BibitemShut {NoStop}%
\bibitem [{\citenamefont {Philips}\ \emph {et~al.}(2022)\citenamefont
  {Philips}, \citenamefont {Mądzik}, \citenamefont {Amitonov}, \citenamefont
  {de~Snoo}, \citenamefont {Russ}, \citenamefont {Kalhor}, \citenamefont
  {Volk}, \citenamefont {Lawrie}, \citenamefont {Brousse}, \citenamefont
  {Tryputen}, \citenamefont {Wuetz}, \citenamefont {Sammak}, \citenamefont
  {Veldhorst},\ and\ \citenamefont {Vandersypen}}]{Philips2022}%
  \BibitemOpen
  \bibfield  {author} {\bibinfo {author} {\bibfnamefont {S.~G.~J.}\
  \bibnamefont {Philips}}, \bibinfo {author} {\bibfnamefont {M.~T.}\
  \bibnamefont {Mądzik}}, \bibinfo {author} {\bibfnamefont {S.}~\bibnamefont
  {Amitonov}}, \bibinfo {author} {\bibfnamefont {S.~L.}\ \bibnamefont
  {de~Snoo}}, \bibinfo {author} {\bibfnamefont {M.}~\bibnamefont {Russ}},
  \bibinfo {author} {\bibfnamefont {N.}~\bibnamefont {Kalhor}}, \bibinfo
  {author} {\bibfnamefont {C.}~\bibnamefont {Volk}}, \bibinfo {author}
  {\bibfnamefont {W.~I.~L.}\ \bibnamefont {Lawrie}}, \bibinfo {author}
  {\bibfnamefont {D.}~\bibnamefont {Brousse}}, \bibinfo {author} {\bibfnamefont
  {L.}~\bibnamefont {Tryputen}}, \bibinfo {author} {\bibfnamefont {B.~P.}\
  \bibnamefont {Wuetz}}, \bibinfo {author} {\bibfnamefont {A.}~\bibnamefont
  {Sammak}}, \bibinfo {author} {\bibfnamefont {G.}~\bibnamefont {Veldhorst},
  \bibfnamefont {M.and~Scappucci}},\ and\ \bibinfo {author} {\bibfnamefont
  {L.~M.~K.}\ \bibnamefont {Vandersypen}},\ }\bibfield  {title} {\bibinfo
  {title} {Universal control of a six-qubit quantum processor in silicon},\
  }\href@noop {} {\bibfield  {journal} {\bibinfo  {journal} {Nature}\ }\textbf
  {\bibinfo {volume} {609}},\ \bibinfo {pages} {919} (\bibinfo {year}
  {2022})}\BibitemShut {NoStop}%
\bibitem [{\citenamefont {Borsoi}\ \emph {et~al.}(2023)\citenamefont {Borsoi},
  \citenamefont {Hendrickx}, \citenamefont {John}, \citenamefont {Meyer},
  \citenamefont {Motz}, \citenamefont {van Riggelen}, \citenamefont {Sammak},
  \citenamefont {de~Snoo}, \citenamefont {Scappucci},\ and\ \citenamefont
  {Veldhorst}}]{Borsoi2023}%
  \BibitemOpen
  \bibfield  {author} {\bibinfo {author} {\bibfnamefont {F.}~\bibnamefont
  {Borsoi}}, \bibinfo {author} {\bibfnamefont {N.~W.}\ \bibnamefont
  {Hendrickx}}, \bibinfo {author} {\bibfnamefont {V.}~\bibnamefont {John}},
  \bibinfo {author} {\bibfnamefont {M.}~\bibnamefont {Meyer}}, \bibinfo
  {author} {\bibfnamefont {S.}~\bibnamefont {Motz}}, \bibinfo {author}
  {\bibfnamefont {F.}~\bibnamefont {van Riggelen}}, \bibinfo {author}
  {\bibfnamefont {A.}~\bibnamefont {Sammak}}, \bibinfo {author} {\bibfnamefont
  {S.~L.}\ \bibnamefont {de~Snoo}}, \bibinfo {author} {\bibfnamefont
  {G.}~\bibnamefont {Scappucci}},\ and\ \bibinfo {author} {\bibfnamefont
  {M.}~\bibnamefont {Veldhorst}},\ }\bibfield  {title} {\bibinfo {title}
  {Shared control of a 16 semiconductor quantum dot crossbar array},\
  }\href@noop {} {\bibfield  {journal} {\bibinfo  {journal} {Nature
  Nanotechnology}\ } (\bibinfo {year} {2023})}\BibitemShut {NoStop}%
\bibitem [{\citenamefont {Fowler}\ \emph {et~al.}(2012)\citenamefont {Fowler},
  \citenamefont {Mariantoni}, \citenamefont {Martinis},\ and\ \citenamefont
  {Cleland}}]{Fowler2012}%
  \BibitemOpen
  \bibfield  {author} {\bibinfo {author} {\bibfnamefont {A.~G.}\ \bibnamefont
  {Fowler}}, \bibinfo {author} {\bibfnamefont {M.}~\bibnamefont {Mariantoni}},
  \bibinfo {author} {\bibfnamefont {J.~M.}\ \bibnamefont {Martinis}},\ and\
  \bibinfo {author} {\bibfnamefont {A.~N.}\ \bibnamefont {Cleland}},\
  }\bibfield  {title} {\bibinfo {title} {Surface codes: Towards practical
  large-scale quantum computation},\ }\href@noop {} {\bibfield  {journal}
  {\bibinfo  {journal} {Physical Review A}\ }\textbf {\bibinfo {volume} {86}},\
  \bibinfo {pages} {032324} (\bibinfo {year} {2012})}\BibitemShut {NoStop}%
\bibitem [{\citenamefont {Wecker}\ \emph {et~al.}(2014)\citenamefont {Wecker},
  \citenamefont {Bauer}, \citenamefont {Clark}, \citenamefont {Hastings},\ and\
  \citenamefont {Troyer}}]{Wecker2014}%
  \BibitemOpen
  \bibfield  {author} {\bibinfo {author} {\bibfnamefont {D.}~\bibnamefont
  {Wecker}}, \bibinfo {author} {\bibfnamefont {B.}~\bibnamefont {Bauer}},
  \bibinfo {author} {\bibfnamefont {B.~K.}\ \bibnamefont {Clark}}, \bibinfo
  {author} {\bibfnamefont {M.~B.}\ \bibnamefont {Hastings}},\ and\ \bibinfo
  {author} {\bibfnamefont {M.}~\bibnamefont {Troyer}},\ }\bibfield  {title}
  {\bibinfo {title} {Gate-count estimates for performing quantum chemistry on
  small quantum computers},\ }\href@noop {} {\bibfield  {journal} {\bibinfo
  {journal} {Physical Review A}\ }\textbf {\bibinfo {volume} {90}},\ \bibinfo
  {pages} {022305} (\bibinfo {year} {2014})}\BibitemShut {NoStop}%
\bibitem [{\citenamefont {Terhal}(2015)}]{Terhal2015}%
  \BibitemOpen
  \bibfield  {author} {\bibinfo {author} {\bibfnamefont {B.~M.}\ \bibnamefont
  {Terhal}},\ }\bibfield  {title} {\bibinfo {title} {Quantum error correction
  for quantum memories},\ }\href@noop {} {\bibfield  {journal} {\bibinfo
  {journal} {Reviews of Modern Physics}\ }\textbf {\bibinfo {volume} {87}},\
  \bibinfo {pages} {307} (\bibinfo {year} {2015})}\BibitemShut {NoStop}%
\bibitem [{\citenamefont {Loss}\ and\ \citenamefont
  {DiVincenzo}(1998)}]{LossDiVincenzo1998}%
  \BibitemOpen
  \bibfield  {author} {\bibinfo {author} {\bibfnamefont {D.}~\bibnamefont
  {Loss}}\ and\ \bibinfo {author} {\bibfnamefont {D.~P.}\ \bibnamefont
  {DiVincenzo}},\ }\bibfield  {title} {\bibinfo {title} {Quantum computation
  with quantum dots},\ }\href@noop {} {\bibfield  {journal} {\bibinfo
  {journal} {Physical Review A}\ }\textbf {\bibinfo {volume} {57}},\ \bibinfo
  {pages} {120} (\bibinfo {year} {1998})}\BibitemShut {NoStop}%
\bibitem [{\citenamefont {Schäffler}(1997)}]{Schäffler1997}%
  \BibitemOpen
  \bibfield  {author} {\bibinfo {author} {\bibfnamefont {F.}~\bibnamefont
  {Schäffler}},\ }\bibfield  {title} {\bibinfo {title} {High-mobility {S}i and
  {G}e structures},\ }\href@noop {} {\bibfield  {journal} {\bibinfo  {journal}
  {Semiconductor Science and Technology}\ }\textbf {\bibinfo {volume} {12}},\
  \bibinfo {pages} {1515} (\bibinfo {year} {1997})}\BibitemShut {NoStop}%
\bibitem [{\citenamefont {Borselli}\ \emph {et~al.}(2011)\citenamefont
  {Borselli}, \citenamefont {Eng}, \citenamefont {Croke}, \citenamefont
  {Maune}, \citenamefont {Huang}, \citenamefont {Ross}, \citenamefont
  {Kiselev}, \citenamefont {Deelman}, \citenamefont {Alvarado-Rodriguez},
  \citenamefont {Schmitz}, \citenamefont {Sokolich}, \citenamefont {Holabird},
  \citenamefont {Hazard}, \citenamefont {Gyure},\ and\ \citenamefont
  {Hunter}}]{Borselli2011}%
  \BibitemOpen
  \bibfield  {author} {\bibinfo {author} {\bibfnamefont {M.~G.}\ \bibnamefont
  {Borselli}}, \bibinfo {author} {\bibfnamefont {K.}~\bibnamefont {Eng}},
  \bibinfo {author} {\bibfnamefont {E.~T.}\ \bibnamefont {Croke}}, \bibinfo
  {author} {\bibfnamefont {B.~M.}\ \bibnamefont {Maune}}, \bibinfo {author}
  {\bibfnamefont {B.}~\bibnamefont {Huang}}, \bibinfo {author} {\bibfnamefont
  {R.~S.}\ \bibnamefont {Ross}}, \bibinfo {author} {\bibfnamefont {A.~A.}\
  \bibnamefont {Kiselev}}, \bibinfo {author} {\bibfnamefont {P.~W.}\
  \bibnamefont {Deelman}}, \bibinfo {author} {\bibfnamefont {I.}~\bibnamefont
  {Alvarado-Rodriguez}}, \bibinfo {author} {\bibfnamefont {A.~E.}\ \bibnamefont
  {Schmitz}}, \bibinfo {author} {\bibfnamefont {M.}~\bibnamefont {Sokolich}},
  \bibinfo {author} {\bibfnamefont {K.~S.}\ \bibnamefont {Holabird}}, \bibinfo
  {author} {\bibfnamefont {T.~M.}\ \bibnamefont {Hazard}}, \bibinfo {author}
  {\bibfnamefont {M.~F.}\ \bibnamefont {Gyure}},\ and\ \bibinfo {author}
  {\bibfnamefont {A.~T.}\ \bibnamefont {Hunter}},\ }\bibfield  {title}
  {\bibinfo {title} {Pauli spin blockade in undoped {S}i/{S}i{G}e two-electron
  double quantum dots},\ }\href@noop {} {\bibfield  {journal} {\bibinfo
  {journal} {Applied Physics Letters}\ }\textbf {\bibinfo {volume} {99}},\
  \bibinfo {pages} {063109} (\bibinfo {year} {2011})}\BibitemShut {NoStop}%
\bibitem [{\citenamefont {Mi}\ \emph {et~al.}(2015)\citenamefont {Mi},
  \citenamefont {Hazard}, \citenamefont {Payette}, \citenamefont {Wang},
  \citenamefont {Zajac}, \citenamefont {Cady},\ and\ \citenamefont
  {Petta}}]{Mi2015}%
  \BibitemOpen
  \bibfield  {author} {\bibinfo {author} {\bibfnamefont {X.}~\bibnamefont
  {Mi}}, \bibinfo {author} {\bibfnamefont {T.~M.}\ \bibnamefont {Hazard}},
  \bibinfo {author} {\bibfnamefont {C.}~\bibnamefont {Payette}}, \bibinfo
  {author} {\bibfnamefont {K.}~\bibnamefont {Wang}}, \bibinfo {author}
  {\bibfnamefont {D.~M.}\ \bibnamefont {Zajac}}, \bibinfo {author}
  {\bibfnamefont {J.~V.}\ \bibnamefont {Cady}},\ and\ \bibinfo {author}
  {\bibfnamefont {J.~R.}\ \bibnamefont {Petta}},\ }\bibfield  {title} {\bibinfo
  {title} {Magnetotransport studies of mobility limiting mechanisms in undoped
  {S}i/{S}i{G}e heterostructures},\ }\href@noop {} {\bibfield  {journal}
  {\bibinfo  {journal} {Physical Review B}\ }\textbf {\bibinfo {volume} {92}},\
  \bibinfo {pages} {035304} (\bibinfo {year} {2015})}\BibitemShut {NoStop}%
\bibitem [{\citenamefont {Li}\ \emph {et~al.}(2015)\citenamefont {Li},
  \citenamefont {Sookchoo}, \citenamefont {Cui}, \citenamefont {Mohr},
  \citenamefont {Savage}, \citenamefont {Foote}, \citenamefont {Jacobson},
  \citenamefont {Sánchez-Pérez}, \citenamefont {Paskiewicz}, \citenamefont
  {Wu}, \citenamefont {Ward}, \citenamefont {Coppersmith}, \citenamefont
  {Eriksson},\ and\ \citenamefont {Lagally}}]{Li2015}%
  \BibitemOpen
  \bibfield  {author} {\bibinfo {author} {\bibfnamefont {Y.~S.}\ \bibnamefont
  {Li}}, \bibinfo {author} {\bibfnamefont {P.}~\bibnamefont {Sookchoo}},
  \bibinfo {author} {\bibfnamefont {X.}~\bibnamefont {Cui}}, \bibinfo {author}
  {\bibfnamefont {R.}~\bibnamefont {Mohr}}, \bibinfo {author} {\bibfnamefont
  {D.~E.}\ \bibnamefont {Savage}}, \bibinfo {author} {\bibfnamefont {R.~H.}\
  \bibnamefont {Foote}}, \bibinfo {author} {\bibfnamefont {R.~B.}\ \bibnamefont
  {Jacobson}}, \bibinfo {author} {\bibfnamefont {J.~R.}\ \bibnamefont
  {Sánchez-Pérez}}, \bibinfo {author} {\bibfnamefont {D.~M.}\ \bibnamefont
  {Paskiewicz}}, \bibinfo {author} {\bibfnamefont {X.}~\bibnamefont {Wu}},
  \bibinfo {author} {\bibfnamefont {D.~R.}\ \bibnamefont {Ward}}, \bibinfo
  {author} {\bibfnamefont {S.~N.}\ \bibnamefont {Coppersmith}}, \bibinfo
  {author} {\bibfnamefont {M.~A.}\ \bibnamefont {Eriksson}},\ and\ \bibinfo
  {author} {\bibfnamefont {M.~G.}\ \bibnamefont {Lagally}},\ }\bibfield
  {title} {\bibinfo {title} {Electronic transport properties of epitaxial
  {S}i/{S}i{G}e heterostructures grown on single-crystal {S}i{G}e
  nanomembranes},\ }\href@noop {} {\bibfield  {journal} {\bibinfo  {journal}
  {ACS Nano}\ }\textbf {\bibinfo {volume} {9}},\ \bibinfo {pages} {4891}
  (\bibinfo {year} {2015})}\BibitemShut {NoStop}%
\bibitem [{\citenamefont {Degli~Esposti}\ \emph {et~al.}(2022)\citenamefont
  {Degli~Esposti}, \citenamefont {Paquelet~Wuetz}, \citenamefont {Fezzi},
  \citenamefont {Lodari}, \citenamefont {Sammak},\ and\ \citenamefont
  {Scappucci}}]{Esposti2022}%
  \BibitemOpen
  \bibfield  {author} {\bibinfo {author} {\bibfnamefont {D.}~\bibnamefont
  {Degli~Esposti}}, \bibinfo {author} {\bibfnamefont {B.}~\bibnamefont
  {Paquelet~Wuetz}}, \bibinfo {author} {\bibfnamefont {V.}~\bibnamefont
  {Fezzi}}, \bibinfo {author} {\bibfnamefont {M.}~\bibnamefont {Lodari}},
  \bibinfo {author} {\bibfnamefont {A.}~\bibnamefont {Sammak}},\ and\ \bibinfo
  {author} {\bibfnamefont {G.}~\bibnamefont {Scappucci}},\ }\bibfield  {title}
  {\bibinfo {title} {Wafer-scale low-disorder {2DEG} in $^{28}${S}i/{S}i{G}e
  without an epitaxial {S}i cap},\ }\href@noop {} {\bibfield  {journal}
  {\bibinfo  {journal} {Applied Physics Letters}\ }\textbf {\bibinfo {volume}
  {120}},\ \bibinfo {pages} {184003} (\bibinfo {year} {2022})}\BibitemShut
  {NoStop}%
\bibitem [{\citenamefont {Paquelet~Wuetz}\ \emph {et~al.}(2023)\citenamefont
  {Paquelet~Wuetz}, \citenamefont {Degli~Esposti}, \citenamefont {Zwerver},
  \citenamefont {Amitonov}, \citenamefont {Botifoll}, \citenamefont {Arbiol},
  \citenamefont {Sammak}, \citenamefont {Vandersypen}, \citenamefont {Russ},\
  and\ \citenamefont {Scappucci}}]{Wuetz2023}%
  \BibitemOpen
  \bibfield  {author} {\bibinfo {author} {\bibfnamefont {B.}~\bibnamefont
  {Paquelet~Wuetz}}, \bibinfo {author} {\bibfnamefont {D.}~\bibnamefont
  {Degli~Esposti}}, \bibinfo {author} {\bibfnamefont {A.-M.~J.}\ \bibnamefont
  {Zwerver}}, \bibinfo {author} {\bibfnamefont {S.~V.}\ \bibnamefont
  {Amitonov}}, \bibinfo {author} {\bibfnamefont {M.}~\bibnamefont {Botifoll}},
  \bibinfo {author} {\bibfnamefont {J.}~\bibnamefont {Arbiol}}, \bibinfo
  {author} {\bibfnamefont {A.}~\bibnamefont {Sammak}}, \bibinfo {author}
  {\bibfnamefont {L.~M.~K.}\ \bibnamefont {Vandersypen}}, \bibinfo {author}
  {\bibfnamefont {M.}~\bibnamefont {Russ}},\ and\ \bibinfo {author}
  {\bibfnamefont {G.}~\bibnamefont {Scappucci}},\ }\bibfield  {title} {\bibinfo
  {title} {Reducing charge noise in quantum dots by using thin silicon quantum
  wells},\ }\href@noop {} {\bibfield  {journal} {\bibinfo  {journal} {Nature
  Communications}\ }\textbf {\bibinfo {volume} {14}},\ \bibinfo {pages} {1385}
  (\bibinfo {year} {2023})}\BibitemShut {NoStop}%
\bibitem [{\citenamefont {Stehouwer}\ \emph {et~al.}(2023)\citenamefont
  {Stehouwer}, \citenamefont {Tosato}, \citenamefont {Degli~Esposti},
  \citenamefont {Costa}, \citenamefont {Veldhorst}, \citenamefont {Sammak},\
  and\ \citenamefont {Scappucci}}]{Stehouwer2023}%
  \BibitemOpen
  \bibfield  {author} {\bibinfo {author} {\bibfnamefont {L.~E.~A.}\
  \bibnamefont {Stehouwer}}, \bibinfo {author} {\bibfnamefont {A.}~\bibnamefont
  {Tosato}}, \bibinfo {author} {\bibfnamefont {D.}~\bibnamefont
  {Degli~Esposti}}, \bibinfo {author} {\bibfnamefont {D.}~\bibnamefont
  {Costa}}, \bibinfo {author} {\bibfnamefont {M.}~\bibnamefont {Veldhorst}},
  \bibinfo {author} {\bibfnamefont {A.}~\bibnamefont {Sammak}},\ and\ \bibinfo
  {author} {\bibfnamefont {G.}~\bibnamefont {Scappucci}},\ }\bibfield  {title}
  {\bibinfo {title} {Germanium wafers for strained quantum wells with low
  disorder},\ }\href@noop {} {\bibfield  {journal} {\bibinfo  {journal}
  {Applied Physics Letters}\ }\textbf {\bibinfo {volume} {123}} (\bibinfo
  {year} {2023})}\BibitemShut {NoStop}%
\bibitem [{\citenamefont {Myronov}\ \emph {et~al.}(2023)\citenamefont
  {Myronov}, \citenamefont {Kycia}, \citenamefont {Waldron}, \citenamefont
  {Jiang}, \citenamefont {Bogan}, \citenamefont {Coleridge},\ and\
  \citenamefont {Studenikin}}]{Myronov2023}%
  \BibitemOpen
  \bibfield  {author} {\bibinfo {author} {\bibfnamefont {M.}~\bibnamefont
  {Myronov}}, \bibinfo {author} {\bibfnamefont {J.}~\bibnamefont {Kycia}},
  \bibinfo {author} {\bibfnamefont {P.}~\bibnamefont {Waldron}}, \bibinfo
  {author} {\bibfnamefont {P.}~\bibnamefont {Jiang}, \bibfnamefont
  {W.and~Barrios}}, \bibinfo {author} {\bibfnamefont {A.}~\bibnamefont
  {Bogan}}, \bibinfo {author} {\bibfnamefont {P.}~\bibnamefont {Coleridge}},\
  and\ \bibinfo {author} {\bibfnamefont {S.}~\bibnamefont {Studenikin}},\
  }\bibfield  {title} {\bibinfo {title} {Holes outperform electrons in group
  {IV} semiconductor materials},\ }\href@noop {} {\bibfield  {journal}
  {\bibinfo  {journal} {Small Science}\ }\textbf {\bibinfo {volume} {3}},\
  \bibinfo {pages} {2200094} (\bibinfo {year} {2023})}\BibitemShut {NoStop}%
\bibitem [{\citenamefont {Dodson}\ \emph {et~al.}(2020)\citenamefont {Dodson},
  \citenamefont {Holman}, \citenamefont {Thorgrimsson}, \citenamefont {Neyens},
  \citenamefont {MacQuarrie}, \citenamefont {McJunkin}, \citenamefont {Foote},
  \citenamefont {Edge}, \citenamefont {Coppersmith},\ and\ \citenamefont
  {Eriksson}}]{Dodson2020}%
  \BibitemOpen
  \bibfield  {author} {\bibinfo {author} {\bibfnamefont {J.~P.}\ \bibnamefont
  {Dodson}}, \bibinfo {author} {\bibfnamefont {.}~\bibnamefont {Holman}},
  \bibinfo {author} {\bibfnamefont {B.}~\bibnamefont {Thorgrimsson}}, \bibinfo
  {author} {\bibfnamefont {S.~F.}\ \bibnamefont {Neyens}}, \bibinfo {author}
  {\bibfnamefont {E.~R.}\ \bibnamefont {MacQuarrie}}, \bibinfo {author}
  {\bibfnamefont {T.}~\bibnamefont {McJunkin}}, \bibinfo {author}
  {\bibfnamefont {R.~H.}\ \bibnamefont {Foote}}, \bibinfo {author}
  {\bibfnamefont {L.~F.}\ \bibnamefont {Edge}}, \bibinfo {author}
  {\bibfnamefont {S.~N.}\ \bibnamefont {Coppersmith}},\ and\ \bibinfo {author}
  {\bibfnamefont {M.~A.}\ \bibnamefont {Eriksson}},\ }\bibfield  {title}
  {\bibinfo {title} {Fabrication process and failure analysis for robust
  quantum dots in silicon},\ }\href@noop {} {\bibfield  {journal} {\bibinfo
  {journal} {Nanotechnology}\ }\textbf {\bibinfo {volume} {31}},\ \bibinfo
  {pages} {505001} (\bibinfo {year} {2020})}\BibitemShut {NoStop}%
\bibitem [{\citenamefont {Lawrie}\ \emph {et~al.}(2020)\citenamefont {Lawrie},
  \citenamefont {Eenink}, \citenamefont {Hendrickx}, \citenamefont {Boter},
  \citenamefont {Petit}, \citenamefont {Amitonov}, \citenamefont {Lodari},
  \citenamefont {Paquelet~Wuetz}, \citenamefont {Volk}, \citenamefont
  {Philips}, \citenamefont {Droulers}, \citenamefont {Kalhor}, \citenamefont
  {van Riggelen}, \citenamefont {Brousse}, \citenamefont {Sammak},
  \citenamefont {Vandersypen}, \citenamefont {Scappucci},\ and\ \citenamefont
  {Veldhorst}}]{Lawrie2020}%
  \BibitemOpen
  \bibfield  {author} {\bibinfo {author} {\bibfnamefont {W.~I.~L.}\
  \bibnamefont {Lawrie}}, \bibinfo {author} {\bibfnamefont {H.~G.~J.}\
  \bibnamefont {Eenink}}, \bibinfo {author} {\bibfnamefont {N.~W.}\
  \bibnamefont {Hendrickx}}, \bibinfo {author} {\bibfnamefont {J.~M.}\
  \bibnamefont {Boter}}, \bibinfo {author} {\bibfnamefont {L.}~\bibnamefont
  {Petit}}, \bibinfo {author} {\bibfnamefont {S.~V.}\ \bibnamefont {Amitonov}},
  \bibinfo {author} {\bibfnamefont {M.}~\bibnamefont {Lodari}}, \bibinfo
  {author} {\bibfnamefont {B.}~\bibnamefont {Paquelet~Wuetz}}, \bibinfo
  {author} {\bibfnamefont {C.}~\bibnamefont {Volk}}, \bibinfo {author}
  {\bibfnamefont {S.~G.~J.}\ \bibnamefont {Philips}}, \bibinfo {author}
  {\bibfnamefont {G.}~\bibnamefont {Droulers}}, \bibinfo {author}
  {\bibfnamefont {N.}~\bibnamefont {Kalhor}}, \bibinfo {author} {\bibfnamefont
  {F.}~\bibnamefont {van Riggelen}}, \bibinfo {author} {\bibfnamefont
  {D.}~\bibnamefont {Brousse}}, \bibinfo {author} {\bibfnamefont
  {A.}~\bibnamefont {Sammak}}, \bibinfo {author} {\bibfnamefont {L.~M.~K.}\
  \bibnamefont {Vandersypen}}, \bibinfo {author} {\bibfnamefont
  {G.}~\bibnamefont {Scappucci}},\ and\ \bibinfo {author} {\bibfnamefont
  {M.}~\bibnamefont {Veldhorst}},\ }\bibfield  {title} {\bibinfo {title}
  {Quantum dot arrays in silicon and germanium},\ }\href@noop {} {\bibfield
  {journal} {\bibinfo  {journal} {Applied Physics Letters}\ }\textbf {\bibinfo
  {volume} {116}},\ \bibinfo {pages} {080501} (\bibinfo {year}
  {2020})}\BibitemShut {NoStop}%
\bibitem [{\citenamefont {Ha}\ \emph {et~al.}(2022)\citenamefont {Ha},
  \citenamefont {Ha}, \citenamefont {Choi}, \citenamefont {Tang}, \citenamefont
  {Schmitz}, \citenamefont {Levendorf}, \citenamefont {Lee}, \citenamefont
  {Chappell}, \citenamefont {Adams}, \citenamefont {Hulbert}, \citenamefont
  {Acuna}, \citenamefont {Noah}, \citenamefont {Matten}, \citenamefont {Jura},
  \citenamefont {Wright}, \citenamefont {Rakher},\ and\ \citenamefont
  {Borselli}}]{Ha2022}%
  \BibitemOpen
  \bibfield  {author} {\bibinfo {author} {\bibfnamefont {W.}~\bibnamefont
  {Ha}}, \bibinfo {author} {\bibfnamefont {S.~D.}\ \bibnamefont {Ha}}, \bibinfo
  {author} {\bibfnamefont {M.~D.}\ \bibnamefont {Choi}}, \bibinfo {author}
  {\bibfnamefont {Y.}~\bibnamefont {Tang}}, \bibinfo {author} {\bibfnamefont
  {A.~E.}\ \bibnamefont {Schmitz}}, \bibinfo {author} {\bibfnamefont {M.~P.}\
  \bibnamefont {Levendorf}}, \bibinfo {author} {\bibfnamefont {K.}~\bibnamefont
  {Lee}}, \bibinfo {author} {\bibfnamefont {J.~M.}\ \bibnamefont {Chappell}},
  \bibinfo {author} {\bibfnamefont {T.~S.}\ \bibnamefont {Adams}}, \bibinfo
  {author} {\bibfnamefont {D.~R.}\ \bibnamefont {Hulbert}}, \bibinfo {author}
  {\bibfnamefont {E.}~\bibnamefont {Acuna}}, \bibinfo {author} {\bibfnamefont
  {R.~S.}\ \bibnamefont {Noah}}, \bibinfo {author} {\bibfnamefont {J.~W.}\
  \bibnamefont {Matten}}, \bibinfo {author} {\bibfnamefont {M.~P.}\
  \bibnamefont {Jura}}, \bibinfo {author} {\bibfnamefont {J.~A.}\ \bibnamefont
  {Wright}}, \bibinfo {author} {\bibfnamefont {M.~T.}\ \bibnamefont {Rakher}},\
  and\ \bibinfo {author} {\bibfnamefont {M.~G.}\ \bibnamefont {Borselli}},\
  }\bibfield  {title} {\bibinfo {title} {A flexible design platform for
  {S}i/{S}i{G}e exchange-only qubits with low disorder},\ }\href@noop {}
  {\bibfield  {journal} {\bibinfo  {journal} {Nano Letters}\ }\textbf {\bibinfo
  {volume} {22}},\ \bibinfo {pages} {1443} (\bibinfo {year}
  {2022})}\BibitemShut {NoStop}%
\bibitem [{\citenamefont {Thorbeck}\ and\ \citenamefont
  {Zimmerman}(2015)}]{Thorbeck2015}%
  \BibitemOpen
  \bibfield  {author} {\bibinfo {author} {\bibfnamefont {T.}~\bibnamefont
  {Thorbeck}}\ and\ \bibinfo {author} {\bibfnamefont {N.~M.}\ \bibnamefont
  {Zimmerman}},\ }\bibfield  {title} {\bibinfo {title} {Formation of
  strain-induced quantum dots in gated semiconductor nanostructures},\
  }\href@noop {} {\bibfield  {journal} {\bibinfo  {journal} {AIP Advances}\
  }\textbf {\bibinfo {volume} {5}},\ \bibinfo {pages} {087107} (\bibinfo {year}
  {2015})}\BibitemShut {NoStop}%
\bibitem [{\citenamefont {Park}\ \emph {et~al.}(2016)\citenamefont {Park},
  \citenamefont {Ahn}, \citenamefont {Tilka}, \citenamefont {Sampson},
  \citenamefont {Savage}, \citenamefont {Prance}, \citenamefont {Simmons},
  \citenamefont {Lagally}, \citenamefont {Coppersmith}, \citenamefont
  {Eriksson}, \citenamefont {Holt},\ and\ \citenamefont {Evans}}]{Park2016}%
  \BibitemOpen
  \bibfield  {author} {\bibinfo {author} {\bibfnamefont {J.}~\bibnamefont
  {Park}}, \bibinfo {author} {\bibfnamefont {Y.}~\bibnamefont {Ahn}}, \bibinfo
  {author} {\bibfnamefont {J.~A.}\ \bibnamefont {Tilka}}, \bibinfo {author}
  {\bibfnamefont {K.~C.}\ \bibnamefont {Sampson}}, \bibinfo {author}
  {\bibfnamefont {D.~E.}\ \bibnamefont {Savage}}, \bibinfo {author}
  {\bibfnamefont {J.~R.}\ \bibnamefont {Prance}}, \bibinfo {author}
  {\bibfnamefont {C.~B.}\ \bibnamefont {Simmons}}, \bibinfo {author}
  {\bibfnamefont {M.~G.}\ \bibnamefont {Lagally}}, \bibinfo {author}
  {\bibfnamefont {S.~N.}\ \bibnamefont {Coppersmith}}, \bibinfo {author}
  {\bibfnamefont {M.~A.}\ \bibnamefont {Eriksson}}, \bibinfo {author}
  {\bibfnamefont {M.~V.}\ \bibnamefont {Holt}},\ and\ \bibinfo {author}
  {\bibfnamefont {P.~G.}\ \bibnamefont {Evans}},\ }\bibfield  {title} {\bibinfo
  {title} {Electrode-stress-induced nanoscale disorder in {S}i quantum
  electronic devices},\ }\href@noop {} {\bibfield  {journal} {\bibinfo
  {journal} {APL Materials}\ }\textbf {\bibinfo {volume} {4}},\ \bibinfo
  {pages} {066102} (\bibinfo {year} {2016})}\BibitemShut {NoStop}%
\bibitem [{\citenamefont {Stein}\ \emph {et~al.}(2021)\citenamefont {Stein},
  \citenamefont {Barcikowski}, \citenamefont {Pookpanratana}, \citenamefont
  {Pomeroy},\ and\ \citenamefont {Stewart}}]{Stein2021}%
  \BibitemOpen
  \bibfield  {author} {\bibinfo {author} {\bibfnamefont {R.~M.}\ \bibnamefont
  {Stein}}, \bibinfo {author} {\bibfnamefont {Z.~S.}\ \bibnamefont
  {Barcikowski}}, \bibinfo {author} {\bibfnamefont {S.~J.}\ \bibnamefont
  {Pookpanratana}}, \bibinfo {author} {\bibfnamefont {J.~M.}\ \bibnamefont
  {Pomeroy}},\ and\ \bibinfo {author} {\bibfnamefont {J.}~\bibnamefont
  {Stewart}, \bibfnamefont {M.~D.}},\ }\bibfield  {title} {\bibinfo {title}
  {Alternatives to aluminum gates for silicon quantum devices: Defects and
  strain},\ }\href@noop {} {\bibfield  {journal} {\bibinfo  {journal} {Journal
  of Applied Physics}\ }\textbf {\bibinfo {volume} {130}},\ \bibinfo {pages}
  {115102} (\bibinfo {year} {2021})}\BibitemShut {NoStop}%
\bibitem [{\citenamefont {Zajac}\ \emph {et~al.}(2016)\citenamefont {Zajac},
  \citenamefont {Hazard}, \citenamefont {Mi}, \citenamefont {Nielsen},\ and\
  \citenamefont {Petta}}]{Zajac2016}%
  \BibitemOpen
  \bibfield  {author} {\bibinfo {author} {\bibfnamefont {D.~M.}\ \bibnamefont
  {Zajac}}, \bibinfo {author} {\bibfnamefont {T.~M.}\ \bibnamefont {Hazard}},
  \bibinfo {author} {\bibfnamefont {X.}~\bibnamefont {Mi}}, \bibinfo {author}
  {\bibfnamefont {E.}~\bibnamefont {Nielsen}},\ and\ \bibinfo {author}
  {\bibfnamefont {J.~R.}\ \bibnamefont {Petta}},\ }\bibfield  {title} {\bibinfo
  {title} {Scalable gate architecture for a one-dimensional array of
  semiconductor spin qubits},\ }\href@noop {} {\bibfield  {journal} {\bibinfo
  {journal} {Physical Review Applied}\ }\textbf {\bibinfo {volume} {6}},\
  \bibinfo {pages} {054013} (\bibinfo {year} {2016})}\BibitemShut {NoStop}%
\bibitem [{\citenamefont {Mills}\ \emph {et~al.}(2019)\citenamefont {Mills},
  \citenamefont {Zajac}, \citenamefont {Gullans}, \citenamefont {Schupp},
  \citenamefont {Hazard},\ and\ \citenamefont {Petta}}]{Mills2019}%
  \BibitemOpen
  \bibfield  {author} {\bibinfo {author} {\bibfnamefont {A.~R.}\ \bibnamefont
  {Mills}}, \bibinfo {author} {\bibfnamefont {D.~M.}\ \bibnamefont {Zajac}},
  \bibinfo {author} {\bibfnamefont {M.~J.}\ \bibnamefont {Gullans}}, \bibinfo
  {author} {\bibfnamefont {F.~J.}\ \bibnamefont {Schupp}}, \bibinfo {author}
  {\bibfnamefont {T.~M.}\ \bibnamefont {Hazard}},\ and\ \bibinfo {author}
  {\bibfnamefont {J.~R.}\ \bibnamefont {Petta}},\ }\bibfield  {title} {\bibinfo
  {title} {Shuttling a single charge across a one-dimensional array of silicon
  quantum dots},\ }\href@noop {} {\bibfield  {journal} {\bibinfo  {journal}
  {Nature Communications}\ }\textbf {\bibinfo {volume} {10}},\ \bibinfo {pages}
  {1063} (\bibinfo {year} {2019})}\BibitemShut {NoStop}%
\bibitem [{\citenamefont {Huang}\ \emph {et~al.}(2014)\citenamefont {Huang},
  \citenamefont {Li}, \citenamefont {Chou},\ and\ \citenamefont
  {Sturm}}]{Huang2014}%
  \BibitemOpen
  \bibfield  {author} {\bibinfo {author} {\bibfnamefont {C.-T.}\ \bibnamefont
  {Huang}}, \bibinfo {author} {\bibfnamefont {J.-Y.}\ \bibnamefont {Li}},
  \bibinfo {author} {\bibfnamefont {K.~S.}\ \bibnamefont {Chou}},\ and\
  \bibinfo {author} {\bibfnamefont {J.~C.}\ \bibnamefont {Sturm}},\ }\bibfield
  {title} {\bibinfo {title} {Screening of remote charge scattering sites from
  the oxide/silicon interface of strained {Si} two-dimensional electron gases
  by an intermediate tunable shielding electron layer},\ }\href@noop {}
  {\bibfield  {journal} {\bibinfo  {journal} {Applied Physics Letters}\
  }\textbf {\bibinfo {volume} {104}},\ \bibinfo {pages} {243510} (\bibinfo
  {year} {2014})}\BibitemShut {NoStop}%
\bibitem [{\citenamefont {Laroche}\ \emph {et~al.}(2015)\citenamefont
  {Laroche}, \citenamefont {Huang}, \citenamefont {Nielsen}, \citenamefont
  {Chuang}, \citenamefont {Li}, \citenamefont {Liu},\ and\ \citenamefont
  {Lu}}]{Laroche2015}%
  \BibitemOpen
  \bibfield  {author} {\bibinfo {author} {\bibfnamefont {D.}~\bibnamefont
  {Laroche}}, \bibinfo {author} {\bibfnamefont {S.~H.}\ \bibnamefont {Huang}},
  \bibinfo {author} {\bibfnamefont {E.}~\bibnamefont {Nielsen}}, \bibinfo
  {author} {\bibfnamefont {Y.}~\bibnamefont {Chuang}}, \bibinfo {author}
  {\bibfnamefont {J.~Y.}\ \bibnamefont {Li}}, \bibinfo {author} {\bibfnamefont
  {C.~W.}\ \bibnamefont {Liu}},\ and\ \bibinfo {author} {\bibfnamefont {T.~M.}\
  \bibnamefont {Lu}},\ }\bibfield  {title} {\bibinfo {title} {Scattering
  mechanisms in shallow undoped {S}i/{S}i{G}e quantum wells},\ }\href@noop {}
  {\bibfield  {journal} {\bibinfo  {journal} {AIP Advances}\ }\textbf {\bibinfo
  {volume} {5}},\ \bibinfo {pages} {107106} (\bibinfo {year}
  {2015})}\BibitemShut {NoStop}%
\bibitem [{\citenamefont {Su}\ \emph {et~al.}(2019)\citenamefont {Su},
  \citenamefont {Chou}, \citenamefont {Chuang}, \citenamefont {Lu},\ and\
  \citenamefont {Li}}]{Su2019}%
  \BibitemOpen
  \bibfield  {author} {\bibinfo {author} {\bibfnamefont {Y.-H.}\ \bibnamefont
  {Su}}, \bibinfo {author} {\bibfnamefont {K.-Y.}\ \bibnamefont {Chou}},
  \bibinfo {author} {\bibfnamefont {Y.}~\bibnamefont {Chuang}}, \bibinfo
  {author} {\bibfnamefont {T.-M.}\ \bibnamefont {Lu}},\ and\ \bibinfo {author}
  {\bibfnamefont {J.-Y.}\ \bibnamefont {Li}},\ }\bibfield  {title} {\bibinfo
  {title} {Electron mobility enhancement in an undoped {S}i/{S}i{G}e
  heterostructure by remote carrier screening},\ }\href@noop {} {\bibfield
  {journal} {\bibinfo  {journal} {Journal of Applied Physics}\ }\textbf
  {\bibinfo {volume} {125}},\ \bibinfo {pages} {235705} (\bibinfo {year}
  {2019})}\BibitemShut {NoStop}%
\bibitem [{\citenamefont {Meyer}\ \emph
  {et~al.}(2023{\natexlab{a}})\citenamefont {Meyer}, \citenamefont {D\'eprez},
  \citenamefont {van Abswoude}, \citenamefont {Meijer}, \citenamefont {Liu},
  \citenamefont {Wang}, \citenamefont {Karwal}, \citenamefont {Oosterhout},
  \citenamefont {Borsoi}, \citenamefont {Sammak}, \citenamefont {Hendrickx},
  \citenamefont {Scappucci},\ and\ \citenamefont {Veldhorst}}]{Meyer2023}%
  \BibitemOpen
  \bibfield  {author} {\bibinfo {author} {\bibfnamefont {M.}~\bibnamefont
  {Meyer}}, \bibinfo {author} {\bibfnamefont {C.}~\bibnamefont {D\'eprez}},
  \bibinfo {author} {\bibfnamefont {T.}~\bibnamefont {van Abswoude}}, \bibinfo
  {author} {\bibfnamefont {I.}~\bibnamefont {Meijer}}, \bibinfo {author}
  {\bibfnamefont {D.}~\bibnamefont {Liu}}, \bibinfo {author} {\bibfnamefont
  {C.-A.}\ \bibnamefont {Wang}}, \bibinfo {author} {\bibfnamefont
  {S.}~\bibnamefont {Karwal}}, \bibinfo {author} {\bibfnamefont
  {S.}~\bibnamefont {Oosterhout}}, \bibinfo {author} {\bibfnamefont
  {F.}~\bibnamefont {Borsoi}}, \bibinfo {author} {\bibfnamefont
  {A.}~\bibnamefont {Sammak}}, \bibinfo {author} {\bibfnamefont {N.~W.}\
  \bibnamefont {Hendrickx}}, \bibinfo {author} {\bibfnamefont {G.}~\bibnamefont
  {Scappucci}},\ and\ \bibinfo {author} {\bibfnamefont {M.}~\bibnamefont
  {Veldhorst}},\ }\bibfield  {title} {\bibinfo {title} {Electrical control of
  uniformity in quantum dot devices},\ }\href@noop {} {\bibfield  {journal}
  {\bibinfo  {journal} {Nano Letters}\ }\textbf {\bibinfo {volume} {23}},\
  \bibinfo {pages} {2522} (\bibinfo {year} {2023}{\natexlab{a}})}\BibitemShut
  {NoStop}%
\bibitem [{\citenamefont {Degli~Esposti}\ \emph {et~al.}(2023)\citenamefont
  {Degli~Esposti}, \citenamefont {Stehouwer}, \citenamefont {G\"ul},
  \citenamefont {Samkharadze}, \citenamefont {D\'eprez}, \citenamefont {Meyer},
  \citenamefont {Meijer}, \citenamefont {Tryputen}, \citenamefont {Karwal},
  \citenamefont {Botifoll}, \citenamefont {Arbiol}, \citenamefont {Amitonov},
  \citenamefont {Vandersypen}, \citenamefont {Sammak}, \citenamefont
  {Veldhorst},\ and\ \citenamefont {Scappucci}}]{Esposti2023}%
  \BibitemOpen
  \bibfield  {author} {\bibinfo {author} {\bibfnamefont {D.}~\bibnamefont
  {Degli~Esposti}}, \bibinfo {author} {\bibfnamefont {L.~E.~A.}\ \bibnamefont
  {Stehouwer}}, \bibinfo {author} {\bibfnamefont {O.}~\bibnamefont {G\"ul}},
  \bibinfo {author} {\bibfnamefont {N.}~\bibnamefont {Samkharadze}}, \bibinfo
  {author} {\bibfnamefont {C.}~\bibnamefont {D\'eprez}}, \bibinfo {author}
  {\bibfnamefont {M.}~\bibnamefont {Meyer}}, \bibinfo {author} {\bibfnamefont
  {I.}~\bibnamefont {Meijer}}, \bibinfo {author} {\bibfnamefont
  {L.}~\bibnamefont {Tryputen}}, \bibinfo {author} {\bibfnamefont
  {S.}~\bibnamefont {Karwal}}, \bibinfo {author} {\bibfnamefont
  {M.}~\bibnamefont {Botifoll}}, \bibinfo {author} {\bibfnamefont
  {J.}~\bibnamefont {Arbiol}}, \bibinfo {author} {\bibfnamefont {S.~V.}\
  \bibnamefont {Amitonov}}, \bibinfo {author} {\bibfnamefont {L.~M.~K.}\
  \bibnamefont {Vandersypen}}, \bibinfo {author} {\bibfnamefont
  {A.}~\bibnamefont {Sammak}}, \bibinfo {author} {\bibfnamefont
  {M.}~\bibnamefont {Veldhorst}},\ and\ \bibinfo {author} {\bibfnamefont
  {G.}~\bibnamefont {Scappucci}},\ }\bibfield  {title} {\bibinfo {title} {Low
  disorder and high valley splitting in silicon},\ }\href@noop {} {\bibfield
  {journal} {\bibinfo  {journal} {arXiv}\ ,\ \bibinfo {pages} {2109.07837}}
  (\bibinfo {year} {2023})}\BibitemShut {NoStop}%
\bibitem [{\citenamefont {Unseld}\ \emph {et~al.}(2023)\citenamefont {Unseld},
  \citenamefont {Meyer}, \citenamefont {Madzik}, \citenamefont {Borsoi},
  \citenamefont {de~Snoo}, \citenamefont {Amitonov}, \citenamefont {Sammak},
  \citenamefont {Scappucci}, \citenamefont {Veldhorst},\ and\ \citenamefont
  {Vandersypen}}]{Unseld2023}%
  \BibitemOpen
  \bibfield  {author} {\bibinfo {author} {\bibfnamefont {F.~K.}\ \bibnamefont
  {Unseld}}, \bibinfo {author} {\bibfnamefont {M.}~\bibnamefont {Meyer}},
  \bibinfo {author} {\bibfnamefont {M.~T.}\ \bibnamefont {Madzik}}, \bibinfo
  {author} {\bibfnamefont {F.}~\bibnamefont {Borsoi}}, \bibinfo {author}
  {\bibfnamefont {S.~L.}\ \bibnamefont {de~Snoo}}, \bibinfo {author}
  {\bibfnamefont {S.~V.}\ \bibnamefont {Amitonov}}, \bibinfo {author}
  {\bibfnamefont {A.}~\bibnamefont {Sammak}}, \bibinfo {author} {\bibfnamefont
  {G.}~\bibnamefont {Scappucci}}, \bibinfo {author} {\bibfnamefont
  {M.}~\bibnamefont {Veldhorst}},\ and\ \bibinfo {author} {\bibfnamefont
  {L.~M.~K.}\ \bibnamefont {Vandersypen}},\ }\bibfield  {title} {\bibinfo
  {title} {A 2{D} quantum dot array in planar 28{S}i/{S}i{G}e},\ }\href@noop {}
  {\bibfield  {journal} {\bibinfo  {journal} {Applied Physics Letters}\
  }\textbf {\bibinfo {volume} {123}} (\bibinfo {year} {2023})}\BibitemShut
  {NoStop}%
\bibitem [{\citenamefont {Noiri}\ \emph
  {et~al.}(2022{\natexlab{b}})\citenamefont {Noiri}, \citenamefont {Takeda},
  \citenamefont {Nakajima}, \citenamefont {Kobayashi}, \citenamefont {Sammak},
  \citenamefont {Scappucci},\ and\ \citenamefont {Tarucha}}]{Noiri2022b}%
  \BibitemOpen
  \bibfield  {author} {\bibinfo {author} {\bibfnamefont {A.}~\bibnamefont
  {Noiri}}, \bibinfo {author} {\bibfnamefont {K.}~\bibnamefont {Takeda}},
  \bibinfo {author} {\bibfnamefont {T.}~\bibnamefont {Nakajima}}, \bibinfo
  {author} {\bibfnamefont {T.}~\bibnamefont {Kobayashi}}, \bibinfo {author}
  {\bibfnamefont {A.}~\bibnamefont {Sammak}}, \bibinfo {author} {\bibfnamefont
  {G.}~\bibnamefont {Scappucci}},\ and\ \bibinfo {author} {\bibfnamefont
  {S.}~\bibnamefont {Tarucha}},\ }\bibfield  {title} {\bibinfo {title} {A
  shuttling-based two-qubit logic gate for linking distant silicon quantum
  processors},\ }\href@noop {} {\bibfield  {journal} {\bibinfo  {journal}
  {Nature Communications}\ }\textbf {\bibinfo {volume} {13}},\ \bibinfo {pages}
  {5740} (\bibinfo {year} {2022}{\natexlab{b}})}\BibitemShut {NoStop}%
\bibitem [{\citenamefont {Paquelet~Wuetz}\ \emph {et~al.}(2022)\citenamefont
  {Paquelet~Wuetz}, \citenamefont {Losert}, \citenamefont {Koelling},
  \citenamefont {Stehouwer}, \citenamefont {Zwerver}, \citenamefont {Philips},
  \citenamefont {Madzik}, \citenamefont {Xue}, \citenamefont {Zheng},
  \citenamefont {Lodari}, \citenamefont {Amitonov}, \citenamefont
  {Samkharadze}, \citenamefont {Sammak}, \citenamefont {Vandersypen},
  \citenamefont {Rahman}, \citenamefont {Coppersmith}, \citenamefont
  {Moutanabbir}, \citenamefont {Friesen},\ and\ \citenamefont
  {Scappucci}}]{Wuetz2022}%
  \BibitemOpen
  \bibfield  {author} {\bibinfo {author} {\bibfnamefont {B.}~\bibnamefont
  {Paquelet~Wuetz}}, \bibinfo {author} {\bibfnamefont {M.~P.}\ \bibnamefont
  {Losert}}, \bibinfo {author} {\bibfnamefont {S.}~\bibnamefont {Koelling}},
  \bibinfo {author} {\bibfnamefont {L.~E.~A.}\ \bibnamefont {Stehouwer}},
  \bibinfo {author} {\bibfnamefont {A.-M.~J.}\ \bibnamefont {Zwerver}},
  \bibinfo {author} {\bibfnamefont {S.~G.~J.}\ \bibnamefont {Philips}},
  \bibinfo {author} {\bibfnamefont {M.~T.}\ \bibnamefont {Madzik}}, \bibinfo
  {author} {\bibfnamefont {X.}~\bibnamefont {Xue}}, \bibinfo {author}
  {\bibfnamefont {G.}~\bibnamefont {Zheng}}, \bibinfo {author} {\bibfnamefont
  {M.}~\bibnamefont {Lodari}}, \bibinfo {author} {\bibfnamefont {S.~V.}\
  \bibnamefont {Amitonov}}, \bibinfo {author} {\bibfnamefont {N.}~\bibnamefont
  {Samkharadze}}, \bibinfo {author} {\bibfnamefont {A.}~\bibnamefont {Sammak}},
  \bibinfo {author} {\bibfnamefont {L.~M.~K.}\ \bibnamefont {Vandersypen}},
  \bibinfo {author} {\bibfnamefont {R.}~\bibnamefont {Rahman}}, \bibinfo
  {author} {\bibfnamefont {S.~N.}\ \bibnamefont {Coppersmith}}, \bibinfo
  {author} {\bibfnamefont {O.}~\bibnamefont {Moutanabbir}}, \bibinfo {author}
  {\bibfnamefont {M.}~\bibnamefont {Friesen}},\ and\ \bibinfo {author}
  {\bibfnamefont {G.}~\bibnamefont {Scappucci}},\ }\bibfield  {title} {\bibinfo
  {title} {Atomic fluctuations lifting the energy degeneracy in $\rm{Si/SiGe}$
  quantum dots},\ }\href@noop {} {\bibfield  {journal} {\bibinfo  {journal}
  {Nature Communications}\ }\textbf {\bibinfo {volume} {13}},\ \bibinfo {pages}
  {7730} (\bibinfo {year} {2022})}\BibitemShut {NoStop}%
\bibitem [{\citenamefont {Ziegler}\ \emph {et~al.}(2023)\citenamefont
  {Ziegler}, \citenamefont {Luthi}, \citenamefont {Ramsey}, \citenamefont
  {Borjans}, \citenamefont {Zheng},\ and\ \citenamefont
  {Zwolak}}]{Ziegler2023}%
  \BibitemOpen
  \bibfield  {author} {\bibinfo {author} {\bibfnamefont {J.}~\bibnamefont
  {Ziegler}}, \bibinfo {author} {\bibfnamefont {F.}~\bibnamefont {Luthi}},
  \bibinfo {author} {\bibfnamefont {M.}~\bibnamefont {Ramsey}}, \bibinfo
  {author} {\bibfnamefont {F.}~\bibnamefont {Borjans}}, \bibinfo {author}
  {\bibfnamefont {G.}~\bibnamefont {Zheng}},\ and\ \bibinfo {author}
  {\bibfnamefont {J.~P.}\ \bibnamefont {Zwolak}},\ }\bibfield  {title}
  {\bibinfo {title} {Automated extraction of capacitive coupling for quantum
  dot systems},\ }\href@noop {} {\bibfield  {journal} {\bibinfo  {journal}
  {Physical Review Applied}\ }\textbf {\bibinfo {volume} {19}},\ \bibinfo
  {pages} {054077} (\bibinfo {year} {2023})}\BibitemShut {NoStop}%
\bibitem [{\citenamefont {Li}\ \emph {et~al.}(2018)\citenamefont {Li},
  \citenamefont {Petit}, \citenamefont {Franke}, \citenamefont {Dehollain},
  \citenamefont {Helsen}, \citenamefont {Steudtner}, \citenamefont {Thomas},
  \citenamefont {Yoscovits}, \citenamefont {Singh}, \citenamefont {Wehner},
  \citenamefont {Vandersypen}, \citenamefont {Clarke},\ and\ \citenamefont
  {Veldhorst}}]{Li2018}%
  \BibitemOpen
  \bibfield  {author} {\bibinfo {author} {\bibfnamefont {R.}~\bibnamefont
  {Li}}, \bibinfo {author} {\bibfnamefont {L.}~\bibnamefont {Petit}}, \bibinfo
  {author} {\bibfnamefont {D.~P.}\ \bibnamefont {Franke}}, \bibinfo {author}
  {\bibfnamefont {J.~P.}\ \bibnamefont {Dehollain}}, \bibinfo {author}
  {\bibfnamefont {J.}~\bibnamefont {Helsen}}, \bibinfo {author} {\bibfnamefont
  {M.}~\bibnamefont {Steudtner}}, \bibinfo {author} {\bibfnamefont {N.~K.}\
  \bibnamefont {Thomas}}, \bibinfo {author} {\bibfnamefont {Z.~R.}\
  \bibnamefont {Yoscovits}}, \bibinfo {author} {\bibfnamefont {K.~J.}\
  \bibnamefont {Singh}}, \bibinfo {author} {\bibfnamefont {S.}~\bibnamefont
  {Wehner}}, \bibinfo {author} {\bibfnamefont {L.~M.~K.}\ \bibnamefont
  {Vandersypen}}, \bibinfo {author} {\bibfnamefont {J.}~\bibnamefont
  {Clarke}},\ and\ \bibinfo {author} {\bibfnamefont {M.}~\bibnamefont
  {Veldhorst}},\ }\bibfield  {title} {\bibinfo {title} {A crossbar network for
  silicon quantum dot qubits},\ }\href@noop {} {\bibfield  {journal} {\bibinfo
  {journal} {Science Advances}\ }\textbf {\bibinfo {volume} {4}},\ \bibinfo
  {pages} {eaar3960} (\bibinfo {year} {2018})}\BibitemShut {NoStop}%
\bibitem [{\citenamefont {Hendrickx}\ \emph {et~al.}(2023)\citenamefont
  {Hendrickx}, \citenamefont {Massai}, \citenamefont {Mergenthaler},
  \citenamefont {Schupp}, \citenamefont {Paredes}, \citenamefont {Bedell},
  \citenamefont {Salis},\ and\ \citenamefont {Fuhrer}}]{Hendrickx2023}%
  \BibitemOpen
  \bibfield  {author} {\bibinfo {author} {\bibfnamefont {N.~W.}\ \bibnamefont
  {Hendrickx}}, \bibinfo {author} {\bibfnamefont {L.}~\bibnamefont {Massai}},
  \bibinfo {author} {\bibfnamefont {M.}~\bibnamefont {Mergenthaler}}, \bibinfo
  {author} {\bibfnamefont {F.}~\bibnamefont {Schupp}}, \bibinfo {author}
  {\bibfnamefont {S.}~\bibnamefont {Paredes}}, \bibinfo {author} {\bibfnamefont
  {S.~W.}\ \bibnamefont {Bedell}}, \bibinfo {author} {\bibfnamefont
  {G.}~\bibnamefont {Salis}},\ and\ \bibinfo {author} {\bibfnamefont
  {A.}~\bibnamefont {Fuhrer}},\ }\bibfield  {title} {\bibinfo {title}
  {Sweet-spot operation of a germanium hole spin qubit with highly anisotropic
  noise sensitivity},\ }\href@noop {} {\bibfield  {journal} {\bibinfo
  {journal} {arXiv}\ ,\ \bibinfo {pages} {2305.13150}} (\bibinfo {year}
  {2023})}\BibitemShut {NoStop}%
\bibitem [{\citenamefont {Meyer}\ \emph
  {et~al.}(2023{\natexlab{b}})\citenamefont {Meyer}, \citenamefont {D\'eprez},
  \citenamefont {Meijer}, \citenamefont {Unseld}, \citenamefont {Karwal},
  \citenamefont {Sammak}, \citenamefont {Scappucci}, \citenamefont
  {Vandersypen},\ and\ \citenamefont {Veldhorst}}]{ZENODO_DATA}%
  \BibitemOpen
  \bibfield  {author} {\bibinfo {author} {\bibfnamefont {M.}~\bibnamefont
  {Meyer}}, \bibinfo {author} {\bibfnamefont {C.}~\bibnamefont {D\'eprez}},
  \bibinfo {author} {\bibfnamefont {I.~N.}\ \bibnamefont {Meijer}}, \bibinfo
  {author} {\bibfnamefont {F.~K.}\ \bibnamefont {Unseld}}, \bibinfo {author}
  {\bibfnamefont {S.}~\bibnamefont {Karwal}}, \bibinfo {author} {\bibfnamefont
  {A.}~\bibnamefont {Sammak}}, \bibinfo {author} {\bibfnamefont
  {G.}~\bibnamefont {Scappucci}}, \bibinfo {author} {\bibfnamefont
  {L.}~\bibnamefont {Vandersypen}},\ and\ \bibinfo {author} {\bibfnamefont
  {M.}~\bibnamefont {Veldhorst}},\ }\bibfield  {title} {\bibinfo {title}
  {Dataset underlying the manuscript: Single-electron occupation in quantum dot
  arrays at selectable plunger gate voltage},\ }\href@noop {} {\bibfield
  {journal} {\bibinfo  {journal} {Zenodo.org}\ ,\ \bibinfo {pages}
  {10.5281/zenodo.8322422}} (\bibinfo {year} {2023}{\natexlab{b}})}\BibitemShut
  {NoStop}%
\bibitem [{\citenamefont {Lu}\ \emph {et~al.}(2011)\citenamefont {Lu},
  \citenamefont {Lee}, \citenamefont {Huang}, \citenamefont {Tsui},\ and\
  \citenamefont {Liu}}]{Lu2011}%
  \BibitemOpen
  \bibfield  {author} {\bibinfo {author} {\bibfnamefont {T.~M.}\ \bibnamefont
  {Lu}}, \bibinfo {author} {\bibfnamefont {C.-H.}\ \bibnamefont {Lee}},
  \bibinfo {author} {\bibfnamefont {S.-H.}\ \bibnamefont {Huang}}, \bibinfo
  {author} {\bibfnamefont {D.~C.}\ \bibnamefont {Tsui}},\ and\ \bibinfo
  {author} {\bibfnamefont {C.~W.}\ \bibnamefont {Liu}},\ }\bibfield  {title}
  {\bibinfo {title} {Upper limit of two-dimensional electron density in
  enhancement-mode {S}i/{S}i{G}e heterostructure field-effect transistors},\
  }\href@noop {} {\bibfield  {journal} {\bibinfo  {journal} {Applied Physics
  Letters}\ }\textbf {\bibinfo {volume} {99}},\ \bibinfo {pages} {153510}
  (\bibinfo {year} {2011})}\BibitemShut {NoStop}%
\bibitem [{\citenamefont {Chou}\ \emph {et~al.}(2018)\citenamefont {Chou},
  \citenamefont {Hsu}, \citenamefont {Su}, \citenamefont {Chou}, \citenamefont
  {Chiu}, \citenamefont {Chuang},\ and\ \citenamefont {Li}}]{Chou2018}%
  \BibitemOpen
  \bibfield  {author} {\bibinfo {author} {\bibfnamefont {K.-Y.}\ \bibnamefont
  {Chou}}, \bibinfo {author} {\bibfnamefont {N.-W.}\ \bibnamefont {Hsu}},
  \bibinfo {author} {\bibfnamefont {Y.-H.}\ \bibnamefont {Su}}, \bibinfo
  {author} {\bibfnamefont {C.-T.}\ \bibnamefont {Chou}}, \bibinfo {author}
  {\bibfnamefont {P.-Y.}\ \bibnamefont {Chiu}}, \bibinfo {author}
  {\bibfnamefont {Y.}~\bibnamefont {Chuang}},\ and\ \bibinfo {author}
  {\bibfnamefont {J.-Y.}\ \bibnamefont {Li}},\ }\bibfield  {title} {\bibinfo
  {title} {Temperature dependence of {DC} transport characteristics for a
  two-dimensional electron gas in an undoped {Si/SiGe} heterostructure},\
  }\href@noop {} {\bibfield  {journal} {\bibinfo  {journal} {Applied Physics
  Letters}\ }\textbf {\bibinfo {volume} {112}},\ \bibinfo {pages} {083502}
  (\bibinfo {year} {2018})}\BibitemShut {NoStop}%
\bibitem [{\citenamefont {Connors}\ \emph {et~al.}(2019)\citenamefont
  {Connors}, \citenamefont {Nelson}, \citenamefont {Qiao}, \citenamefont
  {Edge},\ and\ \citenamefont {Nichol}}]{Connors2019}%
  \BibitemOpen
  \bibfield  {author} {\bibinfo {author} {\bibfnamefont {E.~J.}\ \bibnamefont
  {Connors}}, \bibinfo {author} {\bibfnamefont {J.~J.}\ \bibnamefont {Nelson}},
  \bibinfo {author} {\bibfnamefont {H.}~\bibnamefont {Qiao}}, \bibinfo {author}
  {\bibfnamefont {L.~F.}\ \bibnamefont {Edge}},\ and\ \bibinfo {author}
  {\bibfnamefont {J.~M.}\ \bibnamefont {Nichol}},\ }\bibfield  {title}
  {\bibinfo {title} {Low-frequency charge noise in {S}i/{S}i{G}e quantum
  dots},\ }\href@noop {} {\bibfield  {journal} {\bibinfo  {journal} {Physical
  Review B}\ }\textbf {\bibinfo {volume} {100}},\ \bibinfo {pages} {165305}
  (\bibinfo {year} {2019})}\BibitemShut {NoStop}%
\bibitem [{\citenamefont {Struck}\ \emph {et~al.}(2020)\citenamefont {Struck},
  \citenamefont {Hollmann}, \citenamefont {Schauer}, \citenamefont {Fedorets},
  \citenamefont {Schmidbauer}, \citenamefont {Sawano}, \citenamefont {Riemann},
  \citenamefont {Abrosimov}, \citenamefont {Cywinski}, \citenamefont
  {Bougeard},\ and\ \citenamefont {Schreiber}}]{Struck2020}%
  \BibitemOpen
  \bibfield  {author} {\bibinfo {author} {\bibfnamefont {T.}~\bibnamefont
  {Struck}}, \bibinfo {author} {\bibfnamefont {A.}~\bibnamefont {Hollmann}},
  \bibinfo {author} {\bibfnamefont {F.}~\bibnamefont {Schauer}}, \bibinfo
  {author} {\bibfnamefont {O.}~\bibnamefont {Fedorets}}, \bibinfo {author}
  {\bibfnamefont {A.}~\bibnamefont {Schmidbauer}}, \bibinfo {author}
  {\bibfnamefont {.}~\bibnamefont {Sawano}}, \bibinfo {author} {\bibfnamefont
  {H.}~\bibnamefont {Riemann}}, \bibinfo {author} {\bibfnamefont {N.~V.}\
  \bibnamefont {Abrosimov}}, \bibinfo {author} {\bibfnamefont {L.}~\bibnamefont
  {Cywinski}}, \bibinfo {author} {\bibfnamefont {D.}~\bibnamefont {Bougeard}},\
  and\ \bibinfo {author} {\bibfnamefont {L.~R.}\ \bibnamefont {Schreiber}},\
  }\bibfield  {title} {\bibinfo {title} {Low-frequency spin qubit energy
  splitting noise in highly purified $^{28}${S}i/{S}i{G}e},\ }\href@noop {}
  {\bibfield  {journal} {\bibinfo  {journal} {npj Quantum Information}\
  }\textbf {\bibinfo {volume} {6}},\ \bibinfo {pages} {40} (\bibinfo {year}
  {2020})}\BibitemShut {NoStop}%
\bibitem [{\citenamefont {Connors}\ \emph {et~al.}(2022)\citenamefont
  {Connors}, \citenamefont {Nelson}, \citenamefont {Edge},\ and\ \citenamefont
  {Nichol}}]{Connors2022}%
  \BibitemOpen
  \bibfield  {author} {\bibinfo {author} {\bibfnamefont {E.~J.}\ \bibnamefont
  {Connors}}, \bibinfo {author} {\bibfnamefont {J.}~\bibnamefont {Nelson}},
  \bibinfo {author} {\bibfnamefont {L.~F.}\ \bibnamefont {Edge}},\ and\
  \bibinfo {author} {\bibfnamefont {J.~M.}\ \bibnamefont {Nichol}},\ }\bibfield
   {title} {\bibinfo {title} {Charge-noise spectroscopy of {S}i/{S}i{G}e
  quantum dots via dynamically-decoupled exchange oscillations},\ }\href@noop
  {} {\bibfield  {journal} {\bibinfo  {journal} {Nature Communications}\
  }\textbf {\bibinfo {volume} {13}},\ \bibinfo {pages} {940} (\bibinfo {year}
  {2022})}\BibitemShut {NoStop}%
\bibitem [{\citenamefont {Goetzberger}\ \emph {et~al.}(1968)\citenamefont
  {Goetzberger}, \citenamefont {Heine},\ and\ \citenamefont
  {Nicollian}}]{Goetzberger1968}%
  \BibitemOpen
  \bibfield  {author} {\bibinfo {author} {\bibfnamefont {A.}~\bibnamefont
  {Goetzberger}}, \bibinfo {author} {\bibfnamefont {V.}~\bibnamefont {Heine}},\
  and\ \bibinfo {author} {\bibfnamefont {E.~H.}\ \bibnamefont {Nicollian}},\
  }\bibfield  {title} {\bibinfo {title} {Surface states in silicon from charge
  in the oxide coating},\ }\href@noop {} {\bibfield  {journal} {\bibinfo
  {journal} {Applied Physics Letters}\ }\textbf {\bibinfo {volume} {12}},\
  \bibinfo {pages} {95} (\bibinfo {year} {1968})}\BibitemShut {NoStop}%
\bibitem [{\citenamefont {Poindexter}\ and\ \citenamefont
  {Caplan}(1988)}]{Poindexter1988}%
  \BibitemOpen
  \bibfield  {author} {\bibinfo {author} {\bibfnamefont {E.~H.}\ \bibnamefont
  {Poindexter}}\ and\ \bibinfo {author} {\bibfnamefont {P.~J.}\ \bibnamefont
  {Caplan}},\ }\bibfield  {title} {\bibinfo {title} {Electron spin resonance of
  inherent and process induced defects near the {Si/SiO}$_{2}$ interface of
  oxidized silicon wafers},\ }\href@noop {} {\bibfield  {journal} {\bibinfo
  {journal} {Journal of Vacuum Science \& Technology A}\ }\textbf {\bibinfo
  {volume} {6}},\ \bibinfo {pages} {1352} (\bibinfo {year} {1988})}\BibitemShut
  {NoStop}%
\bibitem [{\citenamefont {Lenahan}\ and\ \citenamefont
  {Conley~Jr.}(1998)}]{Lenahan1998}%
  \BibitemOpen
  \bibfield  {author} {\bibinfo {author} {\bibfnamefont {P.~M.}\ \bibnamefont
  {Lenahan}}\ and\ \bibinfo {author} {\bibfnamefont {J.~F.}\ \bibnamefont
  {Conley~Jr.}},\ }\bibfield  {title} {\bibinfo {title} {What can electron
  paramagnetic resonance tell us about the {Si/SiO}$_2$ system?},\ }\href@noop
  {} {\bibfield  {journal} {\bibinfo  {journal} {Journal of Vacuum Science \&
  Technology B: Microelectronics and Nanometer Structures Processing,
  Measurement, and Phenomena}\ }\textbf {\bibinfo {volume} {16}},\ \bibinfo
  {pages} {2134} (\bibinfo {year} {1998})}\BibitemShut {NoStop}%
\bibitem [{\citenamefont {Stesmans}\ \emph {et~al.}(2014)\citenamefont
  {Stesmans}, \citenamefont {Nguyen~Hoang},\ and\ \citenamefont
  {Afanas'ev}}]{Stesmans2014}%
  \BibitemOpen
  \bibfield  {author} {\bibinfo {author} {\bibfnamefont {A.}~\bibnamefont
  {Stesmans}}, \bibinfo {author} {\bibfnamefont {T.}~\bibnamefont
  {Nguyen~Hoang}},\ and\ \bibinfo {author} {\bibfnamefont {V.~V.}\ \bibnamefont
  {Afanas'ev}},\ }\bibfield  {title} {\bibinfo {title} {Hydrogen interaction
  kinetics of {Ge} dangling bonds at the {Si$_{0.25}$Ge$_{0.75}$/SiO$_2$}
  interface},\ }\href@noop {} {\bibfield  {journal} {\bibinfo  {journal}
  {Journal of Applied Physics}\ }\textbf {\bibinfo {volume} {116}},\ \bibinfo
  {pages} {044501} (\bibinfo {year} {2014})}\BibitemShut {NoStop}%
\bibitem [{\citenamefont {Vanheusden}\ \emph {et~al.}(1998)\citenamefont
  {Vanheusden}, \citenamefont {Warren}, \citenamefont {Fleetwood},
  \citenamefont {Schwank}, \citenamefont {Shaneyfelt}, \citenamefont {Draper},
  \citenamefont {Winokur}, \citenamefont {Devine}, \citenamefont {Archer},
  \citenamefont {Brown},\ and\ \citenamefont {Wallace}}]{Vanheusden1998}%
  \BibitemOpen
  \bibfield  {author} {\bibinfo {author} {\bibfnamefont {K.}~\bibnamefont
  {Vanheusden}}, \bibinfo {author} {\bibfnamefont {W.~L.}\ \bibnamefont
  {Warren}}, \bibinfo {author} {\bibfnamefont {D.~M.}\ \bibnamefont
  {Fleetwood}}, \bibinfo {author} {\bibfnamefont {J.~R.}\ \bibnamefont
  {Schwank}}, \bibinfo {author} {\bibfnamefont {M.~R.}\ \bibnamefont
  {Shaneyfelt}}, \bibinfo {author} {\bibfnamefont {B.~L.}\ \bibnamefont
  {Draper}}, \bibinfo {author} {\bibfnamefont {P.~S.}\ \bibnamefont {Winokur}},
  \bibinfo {author} {\bibfnamefont {R.~A.~B.}\ \bibnamefont {Devine}}, \bibinfo
  {author} {\bibfnamefont {L.~B.}\ \bibnamefont {Archer}}, \bibinfo {author}
  {\bibfnamefont {G.~A.}\ \bibnamefont {Brown}},\ and\ \bibinfo {author}
  {\bibfnamefont {R.~M.}\ \bibnamefont {Wallace}},\ }\bibfield  {title}
  {\bibinfo {title} {Chemical kinetics of mobile-proton generation and
  annihilation in {SiO}$_2$ thin films},\ }\href@noop {} {\bibfield  {journal}
  {\bibinfo  {journal} {Applied Physics Letters}\ }\textbf {\bibinfo {volume}
  {73}},\ \bibinfo {pages} {674} (\bibinfo {year} {1998})}\BibitemShut
  {NoStop}%
\bibitem [{\citenamefont {Su}\ \emph {et~al.}(2017)\citenamefont {Su},
  \citenamefont {Chuang}, \citenamefont {Liu}, \citenamefont {Li},\ and\
  \citenamefont {Lu}}]{Su2017}%
  \BibitemOpen
  \bibfield  {author} {\bibinfo {author} {\bibfnamefont {Y.-H.}\ \bibnamefont
  {Su}}, \bibinfo {author} {\bibfnamefont {Y.}~\bibnamefont {Chuang}}, \bibinfo
  {author} {\bibfnamefont {C.-Y.}\ \bibnamefont {Liu}}, \bibinfo {author}
  {\bibfnamefont {J.-Y.}\ \bibnamefont {Li}},\ and\ \bibinfo {author}
  {\bibfnamefont {T.-M.}\ \bibnamefont {Lu}},\ }\bibfield  {title} {\bibinfo
  {title} {Effects of surface tunneling of two-dimensional hole gases in
  undoped {G}e/{G}e{S}i heterostructures},\ }\href@noop {} {\bibfield
  {journal} {\bibinfo  {journal} {Physical Review Materials}\ }\textbf
  {\bibinfo {volume} {1}},\ \bibinfo {pages} {044601} (\bibinfo {year}
  {2017})}\BibitemShut {NoStop}%
\end{thebibliography}%

\end{document}